\documentclass{aa}  

\usepackage{graphicx}
\usepackage{comment}
\graphicspath{{./figures/}}
\usepackage{txfonts}
\usepackage[dvipsnames]{xcolor}
\usepackage{soul}

\usepackage{natbib,twoopt}
\usepackage[breaklinks=true]{hyperref} 
\bibpunct{(}{)}{;}{a}{}{,}             
\makeatletter
  \newcommandtwoopt{\citeads}[3][][]{\href{http://adsabs.harvard.edu/abs/#3}%
    {\def\hyper@linkstart##1##2{}%
     \let\hyper@linkend\@empty\citealp[#1][#2]{#3}}}
  \newcommandtwoopt{\citepads}[3][][]{\href{http://adsabs.harvard.edu/abs/#3}%
    {\def\hyper@linkstart##1##2{}%
     \let\hyper@linkend\@empty\citep[#1][#2]{#3}}}
  \newcommandtwoopt{\citetads}[3][][]{\href{http://adsabs.harvard.edu/abs/#3}%
    {\def\hyper@linkstart##1##2{}%
     \let\hyper@linkend\@empty\citet[#1][#2]{#3}}}
  \newcommandtwoopt{\citeyearads}[3][][]%
    {\href{http://adsabs.harvard.edu/abs/#3}
    {\def\hyper@linkstart##1##2{}%
     \let\hyper@linkend\@empty\citeyear[#1][#2]{#3}}}
\makeatother
\usepackage[]{hyperref}

\begin{document}

   \title{Modeling the mixed-morphology supernova remnant IC~443}

   \subtitle{Origin of its complex morphology and X-ray emission}

   \author{S. Ustamujic
          \inst{1}
          \and
          S. Orlando
          \inst{1}
          \and
          E. Greco
          \inst{2,1}
          \and
          M. Miceli
          \inst{2,1}
          \and
          F. Bocchino
          \inst{1}
          \and
          A. Tutone
          \inst{2,3}
          \and
          G. Peres
          \inst{2,1}        
          }

   \institute{INAF-Osservatorio Astronomico di Palermo, Piazza del Parlamento 1,
             90134 Palermo, Italy\\
              \email{sabina.ustamujic@inaf.it}
         \and
             Dipartimento di Fisica e Chimica E. Segr\`e, Universit\`a di Palermo, 
             Via Archirafi 36, 90123 Palermo, Italy
         \and
             INAF/IASF Palermo, Via Ugo La Malfa 153, I-90146 Palermo, Italy\\
             }

   \date{Received Nov 18, 2020; accepted Mar 4, 2021}
 
  \abstract
   {The morphology and the distribution of material observed in supernova
   remnants (SNRs) reflect the interaction of the supernova (SN) blast 
   wave with the ambient environment, the physical processes associated 
   with the SN explosion and the internal structure of the progenitor 
   star. IC~443 is a mixed-morphology (MM) SNR located in a quite complex 
   environment: it interacts with a molecular cloud in the northwestern 
   and southeastern areas and with an atomic cloud in the northeast.}  
   {In this work we aim at investigating the origin of the complex 
   morphology and multi-thermal X-ray emission observed in SNR~IC~443, 
   through the study of the effect of the inhomogeneous ambient medium in 
   shaping its observed structure, and the exploration of the main 
   parameters characterizing the remnant.}
   {We developed a 3D hydrodynamic (HD) model for IC~443, which describes 
   the interaction of the SNR with the environment, parametrized in agreement 
   with the results of the multi-wavelength data analysis. We performed an 
   ample exploration of the parameter space describing the initial blast 
   wave and the environment, including the mass of the ejecta, the energy 
   and position of the explosion, and the density, structure and geometry of 
   the surrounding clouds. From the simulations, we synthesized the X-ray 
   emission maps and spectra and compared them with actual X-ray data 
   collected by \textit{XMM-Newton}.}
   {Our model explains the origin of the complex X-ray morphology of 
   SNR~IC~443 in a natural way, being able to reproduce, for the first time, 
   most of the observed features, including the centrally-peaked X-ray 
   morphology (characteristic of MM SNRs) when considering the origin of 
   the explosion at the position where the pulsar wind nebula (PWN) 
   CXOU~J061705.3+222127 was at the time of the explosion. In the model 
   which best reproduces the observations, the mass of the ejecta and the 
   energy of the explosion are $\sim 7\,M_\odot$ and $\sim 1\times 10^{51}$~erg, 
   respectively. From the exploration of the parameter space, we found that 
   the density of the clouds is $n > 300$~cm$^{-3}$ and that the age of 
   SNR~IC~443 is $\sim  8000$~yr.}
   {The observed inhomogeneous ambient medium is the main responsible for 
   the complex structure and the X-ray morphology of SNR~IC~443, resulting 
   in a very asymmetric distribution of the ejecta due to the off-centered 
   location of the explosion inside the cavity formed by the clouds. 
   It is argued that the centrally peaked morphology (typical of MMSNRs) 
   is a natural consequence of the interaction with the complex environment. 
   A combination of high resolution X-ray observations and accurate 3D HD 
   modelling is needed to confirm if this scenario is applicable to other 
   MMSNRs.}


   \maketitle
%
\section{Introduction}
\label{sec:intro}

   Supernova remnants (SNRs) are diffuse expanding nebulae that result from 
   a supernova (SN) explosion in which a star ejects violently most of its 
   mass. The ejected material expands from the explosion, interacting with 
   the circumstellar/interstellar medium (CSM/ISM), resulting in a rather 
   complex morphology that reflects the interaction of the SN with the CSM/ISM, 
   but also the asymmetries developed during the SN explosion and the nature 
   of the progenitor star. The SNR IC~443 is one of the best examples of a 
   SNR interacting with a very complex environment, which is made of molecular 
   and atomic clouds \citep{cor77,bur88,rho01,su14}.
   
   IC~443 belongs to the class of mixed-morphology SNR (MMSNR, \citealt{rho98}),
   showing shell-like morphology in the radio band and centrally filled thermal 
   X-ray emission. The majority of MMSNRs are interacting with dense molecular 
   clouds, and nearly half of them have been observed in  $\gamma$-rays 
   \citep{sla15}. The physical origin of the peculiar morphology of MMSNRs is 
   still an open issue. 
   \cite{pet01} stated that this morphology could be due to the nonuniform ambient medium where the SN exploded.
   Traditional models invoke the effects of thermal 
   conduction which induce the evaporation of the dense clouds shocked by the 
   SNRs and the cooling of the ejecta interacting with the clouds (e.g. 
   \citealt{whi91, cox99,she99, zho11, oko20}), though they typically do not 
   consider the role of the ejecta X-ray emission, that is important in many 
   cases \citep{laz06,boc09}. On the other hand, alternative cooling processes 
   suggest that the overionization of the ejecta is due to their rapid free 
   expansion that follows their early heating in case of interaction with 
   dense clouds (e.g. \citealt{mic10, zho11, yam18, gre18}). It is therefore 
   important to carefully model the evolution of MMSNRs, by describing in 
   detail their interaction with the complex environment.
   
   IC~443 (G189.1+3.0) has a diameter of $\sim 50$~arcmin and belongs 
   to the GEM~OB1 association at a distance of 1.5 kpc \citep{pet88, wel03}.
   It appears to consist of two interconnected quasi-spherical subshells of 
   different radii and centroids, that define the usually assumed boundaries 
   of IC~443 \citep{bra86}.
   The remnant age is still uncertain; proposed age for SNR~IC~443 varies 
   from $\sim 3$~kyr \citep{tro08} to $\sim 20-30$~kyr \citep{che99,byk08}.

   Several works have investigated the physical and chemical properties 
   of IC~443 through radio \citep{lea04,lee08,lee12}, infrared 
   \citep{rho01,su14}, X-ray \citep{tro06,tro08,gre18} and $\gamma$-ray 
   \citep{tav10,abd10} observations. IC~443 interacts with a molecular cloud 
   in the northwestern (NW) and southeastern (SE) areas, and with an atomic 
   cloud in the northeast (NE). 
   The dense molecular cloud was first identified by \citet{cor77}, and lies 
   in the foreground of IC~443 forming a semi-toroidal structure 
   \citep{bur88,tro06,su14}. In the NE the remnant is confined by the atomic 
   H~{\footnotesize I} cloud, discovered by \citet{den78}, which is well 
   traced by optical, infrared and very soft X-ray emission \citep[see][]{tro06}.
   The X-ray emission from IC~443 is composed of extended thermal X-ray 
   emission and a number of isolated hard X-ray sources whose emission 
   includes both thermal and non-thermal components 
   \citep[e.g.][]{pet88,boc00,boc03,boc08}. The latter could be a product 
   of the SNR ejecta (or the pulsar wind nebula; PWN) and the surrounding 
   shocked molecular clumps \citep[see][and references therein]{zha18}.
   The gamma-ray emission from IC~443 seems to be associated with the
   interaction of cosmic rays accelerated at the shock front and the nearby 
   molecular clouds \citep{tav10,abd10}. The remnant shows a great diversity 
   and superposition of shocks which may be a natural result of the 
   shock–cloud interactions in a clumpy interstellar medium \citep{sne05,shi11}.
   Indications for the presence of overionized plasma have also been 
   found \citep{yam09,mat17,gre18}, whose origin is still under debate.
   
   The available evidence implies that the SNR shock has encountered a 
   pre-existing high density shell. \cite{tro08} suggested that SNR IC~443 
   has evolved inside a preexisting wind blown bubble, which likely originated 
   from the massive progenitor star of the remnant (probably related to the 
   PWN CXOU~J061705.3+222127). 
   The plerion nebula, discovered by \textit{Chandra} \citep{olb01} and 
   studied by several authors \citep[e.g.][]{boc01,gae06,swa15}, 
   is situated in the southern part of the remnant, 
   but its association with IC~443 is still debated considering its off-center 
   position. Recently, \cite{gre18} detected a Mg-rich jet-like structure 
   in the NW area of IC~443 close to the molecular cloud. Interestingly, 
   the jet emission is mainly due to overionized plasma and its projection 
   towards the remnant interior crosses the position of the neutron star at 
   the time of the explosion of the progenitor star. This strongly suggests 
   that the PWN belongs to IC~443 and that the collimated jet has been 
   produced by the exploding star. 
   
   Here, we investigate the origin of the complex structure and of the 
   multi-thermal X-ray emission of SNR~IC~443 as well as the effects of 
   the inhomogeneous medium in shaping the observed morphology. 
   To this end, we modeled the expansion of the SNR and its interaction with 
   the surrounding environment, parametrized in agreement with the results 
   of the multi-wavelength data analysis. From the simulations we synthesized 
   the thermal X-ray emission and compared it with observations.
   
   The paper is organized as follows. In Sect. 2 we describe the model, 
   the numerical setup, and the synthesis of the thermal X-ray emission; 
   in Sect. 3 we discuss the results; and in Sect. 4 we draw our conclusions.

\section{Hydrodynamic model}
\label{sec:model}

  We developed a three-dimensional (3D) hydrodynamic (HD) model for SNR IC~443, 
  which describes the expansion of the SNR and its interaction with the 
  surrounding CSM/ISM. We followed the evolution of the SNR for 
  $t\approx 10000$~yr by numerically solving the full time-dependent HD 
  equations in a 3D Cartesian coordinate system $(x, y, z)$, including the 
  effects of the radiative losses from optically thin plasma. 
  The HD equations were solved in the conservative form
   \begin{equation}
      \frac{\partial \rho}{\partial t} 
      + \nabla \cdot (\rho \boldsymbol{u}) = 0,
   \end{equation}
   \begin{equation}
      \frac{\partial (\rho \boldsymbol{u}) }{\partial t} + \nabla \cdot 
      (\rho\boldsymbol{u}\boldsymbol{u}) 
      + \nabla P = 0,
   \end{equation}
   \begin{equation}
      \frac{\partial (\rho E) }{\partial t}
      + \nabla \cdot [\boldsymbol{u} (\rho E+P)] = 
      - n_{\mathrm{e}} n_{\mathrm{H}} \Lambda (T),
   \label{eq.energy}
   \end{equation}
   where $E = \epsilon + u^2/2$ is the total gas energy (internal energy 
   $\epsilon$, and kinetic energy) per unit mass, $t$ is the time, 
   $\rho = \mu m_{\mathrm{H}} n_{\mathrm{H}}$ is the mass density, 
   $\mu = 1.29$ is the mean atomic mass of positive ions (assuming cosmic 
   abundances), $m_{\mathrm{H}}$ is the mass of the hydrogen atom, 
   $n_{\mathrm{H}}$ is the hydrogen number density, $\boldsymbol{u}$ is 
   the gas velocity, $T$ is the temperature, and $\Lambda (T)$ represents 
   the optically thin radiative losses per unit emission measure derived with 
   the PINTofALE spectral code \citep{kas00} and with the APED V1.3 atomic 
   line database \citep{smi01}, assuming solar abundances. We used the ideal 
   gas law, $P = (\gamma-1)\rho\epsilon$, where $\gamma=5/3$ is the adiabatic 
   index.

   The calculations were performed using PLUTO \citep{mig07}, a modular 
   Godunov-type code for astrophysical plasmas. The code provides a 
   multiphysics, multialgorithm modular environment particularly oriented 
   towards the treatment of astrophysical high Mach number flows in 
   multiple spatial dimensions. The code was designed to make efficient 
   use of massive parallel computers using the message-passing interface
   (MPI) library for interprocessor communications. The HD equations are
   solved using the HD module available in PLUTO; the integration is 
   performed using the original Piecewise Parabolic Method (PPM)
   reconstruction by \citet[see also \citealt{mil02}]{col84} with a Roe 
   Riemann solver. The adopted scheme is particularly appropriate for 
   describing the shocks formed during the interaction of the remnant 
   with the surrounding inhomogeneous medium, as in our case.
   A monotonized central difference limiter (the least diffusive limiter
   available in PLUTO) for the primitive variables is used.
   PLUTO includes optically thin radiative losses in a fractional step 
   formalism \citep{mig07}, which preserves the second time accuracy, 
   as the advection and source steps are at least second-order accurate;
   the radiative losses ($\Lambda$ values) are computed at the temperature 
   of interest using a table lookup/interpolation method. 
   The code was extended by additional computational modules to evaluate
   the deviations from equilibrium of ionization of the most abundant ions
   (through the computation of the maximum ionization age in each cell 
   of the spatial domain as described in \citealt{orl15}), and the deviations 
   from temperature-equilibration between electrons and ions. 
   For the latter, we included the almost instantaneous heating of 
   electrons at shock fronts up to $kT \sim 0.3$~keV by lower 
   hybrid waves \citep[see][]{gha07}, and the effects of Coulomb collisions 
   for the calculation of ion and electron temperatures in the post-shock 
   plasma \citep[see][for further details]{orl15}.

\subsection{Numerical setup}
\label{sec:num}

   The initial conditions consist of a spherically symmetric distribution 
   of ejecta, representing the remnant of the SN explosion at the age 
   of $\approx 122$~years, which expands through a highly inhomogeneous 
   ISM. At this stage the energy of the SNR is almost entirely kinetic, 
   being the internal energy only a small percentage of the total energy.
   Given the lack of knowledge about the progenitor of SNR IC~443, we explored 
   values in the ranges $4-10\,M_\odot$ for the mass of the ejecta and 
   $1-2.5 \times 10^{51}$~erg for the energy of the explosion.
   The radial density profile of the ejecta is described 
   by two power law segments ($\rho \propto r^{-m}$ on the inside and 
   $\rho \propto r^{-b}$ on the outside), following the density distribution 
   in a core-collapse SN as described by \citet{che05}.
   For our favourite model, we use $m = 1.5$ and $b = 11.3$
   \footnote{This value is consistent 
   with a supernova from a progenitor red supergiant (see Table 4 in 
   \citealt{mat99}).} \citep[see][]{che05}
   leading to an explosion energy of $\sim 1\times 10^{51}$~erg and an 
   ejecta mass of $\sim 7\,M_\odot$. 
   The initial pressure of the ejecta is uniform, while the 
   temperature is the lowest at the center of the initial 
   remnant ($T \approx 20$~K) and the highest in the outermost layers 
   ($T \approx 10^5$~K), following the inverse of the radial profile of density.
   The initial velocity of the ejecta increases linearly with the radial 
   distance from zero, at the center of the remnant, up to 
   $8\times 10^8$~cm~s$^{-1}$ at 1~pc, namely the initial radius of the remnant.
    
\begin{figure}
      \includegraphics*[width=\hsize]{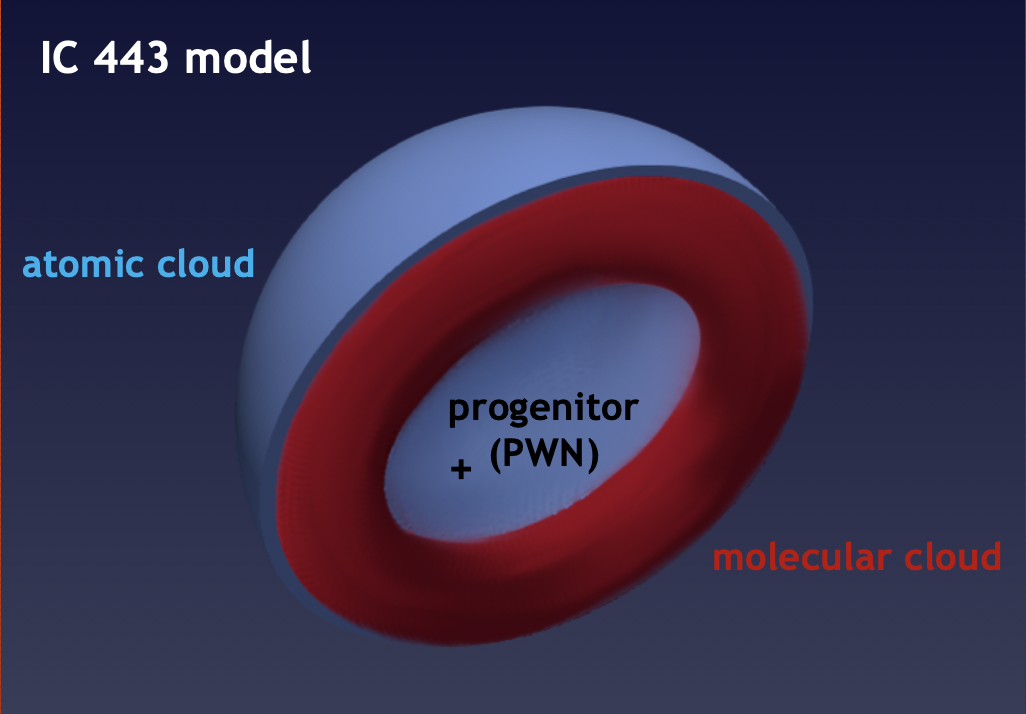}
      \caption{Schematic view of the environment adopted as initial 
      conditions. The toroidal red structure and the spherical blue cap 
      represent the molecular and the atomic clouds respectively. 
      The black cross indicates the position of the PWN.}
      \label{Fig:Initial}
   \end{figure}

   The initial remnant is immersed in a highly inhomogeneous ambient
   environment. We follow the two-shells model proposed by \citet{tro06} to 
   explain the IC~443 morphology \citep[see Fig.~9,][]{tro06}. 
   The medium consists of a toroidal molecular cloud and an spherical cap on 
   the top representing the atomic cloud. 
   In Figure~\ref{Fig:Initial} we present a schematic view of the environment 
   adopted as initial conditions.
   
   The toroidal molecular cloud is azimuthally symmetric about the $z$-axis, 
   centered at the origin of the 3D Cartesian coordinate system 
   $(x_\mathrm{0}, y_\mathrm{0}, z_\mathrm{0}) = (0,0,0)$. 
   The dimensions of the clouds were chosen following the results of
   multi-wavelength data analysis \citep{rho01,tro06,tro08,lee12,su14,gre18}.
   The radius of the torus (i.e., the distance from the origin to the center
   of the tube) is 10~pc, while the radius of the tube is 2.6~pc. The atomic 
   cloud is described as a spherical cap of radius 8~pc centered on the origin, 
   confining the remnant on the positive direction of the $z$-axis.
   
   The clouds play a central role in modifying the expansion of the 
   forward shock and in driving a reverse shock through the ejecta. Since 
   the geometry and density distribution adopted in the paper are idealized, 
   this may introduce some artifacts in the remnant structure if the clouds 
   are assumed to be uniform. On the other side, a clumpy structure of the 
   clouds is expected and often originates from highly compressible turbulence. 
   Thus we adopted a simplistic approach to the treatment of small-scale 
   inhomogeneities in the clouds, since we are not interested in studying 
   in detail the effect of turbulence (as done by \citealt{zha19}). In 
   order to get a non-uniform morphology and density distribution 
   for the clouds, the material is modeled as a set of spherical clumps with 
   radius $\approx$~0.9~pc randomly distributed, filled with spherical sub-clumps
   of radius $\approx$~0.3~pc. For the sake of simplicity, the clouds 
   were defined without considering any velocity fluctuations (which would 
   be anyway much smaller than the shock velocity at the time when the remnant 
   starts to interact with the clouds) and assuming pressure equilibrium 
   among the clumps.
   For the average density we explored values in the range 
   $10<n_\mathrm{a}<10^3$~cm$^{-3}$ and $10^3<n_\mathrm{m}<10^4$~cm$^{-3}$ 
   for the atomic and molecular clouds respectively, in good agreement with 
   the values derived by \citet{rho01} from observations, namely 
   $10 < n \lesssim 1000$~cm$^{-3}$ in the northeastern rim and 
   $n \approx 10^4$~cm$^{-3}$ in the southern molecular ridge.
   The density of the plasma in the spheres follows a normal distribution 
   for both the atomic and the molecular clouds; in our favourite model the 
   mean density is $\approx 300$~cm$^{-3}$ and $\approx 3000$~cm$^{-3}$ 
   respectively. 
   The inter-cloud medium is assumed to be uniform with density $0.2$~cm$^{-3}$ 
   and temperature $10^3$~K.
   The temperatures of the molecular and atomic clouds are set in order to be 
   in pressure equilibrium with the inter-cloud component. 
   The clouds and the inter-cloud medium are defined without considering 
   any bulk velocity, which would be anyway much smaller than the velocity 
   of the forward shock.

   We explored different positions for the explosion (ranging between the 
   geometric center of the cavity formed by the clouds and the very 
   off-centered position of PWN CXOU~J061705.3+222127 suspected to be the 
   compact remnant of the SN) to derive the case that best reproduces the 
   observations. We considered an offset along the $x$-axis, i.e., the origin 
   of the explosion is $(d_\mathrm{x},0,0)$, with $d_\mathrm{x}$ varying in 
   the range $[-5,0]~$pc. The case $d_\mathrm{x}=0$ corresponds to an 
   explosion centered in the cavity formed by the surrounding clouds (see 
   Fig.~\ref{Fig:Initial}). Our favourite model, namely the case that best 
   reproduces the observations, is for $d_\mathrm{x}=-5$~pc, that corresponds 
   to the position of PWN CXOU~J061705.3+222127 at the time of the explosion 
   (as inferred by \citealt{gre18}, taking into account the proper motion of 
   the neutron star).
   
   Table~\ref{parameters} summarizes the parameter space with the 
   corresponding range of values explored, selected according to the values 
   reported in the literature. Given the large number of parameters to be 
   explored, we adopted an educated exploration of the parameter space by 
   starting from a set of initial parameters (e.g. the SN explosion located 
   at the geometric center of the clouds) and by performing an iterative 
   process of trial and error to converge on a set of model parameters that 
   reproduces the main features of X-ray observations of IC~443.
   The values of the model which best reproduce the data are outlined in 
   the last column of Table~\ref{parameters}.
   
   The simulations include passive tracers to follow the evolution of the 
   different plasma components (ejecta - ej - and atomic/molecular clouds; 
   see Fig.~\ref{Fig:Initial}), and to store information on the shocked 
   plasma (time, shock velocity, and shock position, i.e., Lagrangian 
   coordinates, when a cell of the mesh is shocked by either the forward 
   or the reverse shock) required to synthesize the thermal X-ray emission 
   (see Sect.~\ref{sec:synthesis}). The continuity equations of the tracers 
   are solved in addition to our set of HD equations. In the case of tracers 
   associated with the different plasma components, each material is 
   initialized with $C_{\mathrm{i}} = 1$, while $C_{\mathrm{i}} = 0$ 
   elsewhere, where the index $i$ refers to the material in the ejecta, in 
   the atomic cloud and in the molecular cloud. All the other tracers are 
   initialized to zero everywhere.
   
   The computational domain is a Cartesian box extending for $\approx$~$34$~pc 
   in the $x$ and $y$ directions, and for $\approx 30$~pc in the $z$ 
   direction. The box is covered by a uniform grid of $622\times622\times548$ 
   zones, leading to a spatial resolution of $\approx 0.055$~pc. 
   In this way, the initial remnant is covered by $\sim 18$ grid zones. 
   We imposed outflow boundary conditions at all the boundaries.
   
   \begin{table}
      \caption[]{Summary of the initial physical parameters explored: 
      mass of the ejecta, $M_\mathrm{ej}$; energy of the explosion, $E$; 
      offset from the center in the origin of the explosion, $d_\mathrm{x}$;
      density of the atomic, $n_\mathrm{a}$, and molecular, $n_\mathrm{m}$,
      clouds.}
      \label{parameters}
      \centering
      \begin{tabular}{cccc}
      \hline\hline
      Parameters  &  Units  &  Range explored  &  Best model \\
      \hline
      $M_\mathrm{ej}$  &  ($M_\odot$)  &  $[4,10]$  &  $7$ \\
      $E$  &  (erg)  &  $[1,2.5] \times 10^{51}$  &  $1\times 10^{51}$ \\
      $d_\mathrm{x}$  &  (pc)  &  $[-5,0]$  &  $-5$ \\ 
      $n_\mathrm{a}$  &  (cm$^{-3}$)  &  $[10,10^3]$  &  $\sim 300$ \\
      $n_\mathrm{m}$  &  (cm$^{-3}$)  &  $[10^3,10^4]$  &  $\sim 3000$ \\ 
      \hline
      \end{tabular}
   \end{table}

\subsection{Synthesis of X-ray emission}
\label{sec:synthesis}

   We synthesized thermal X-ray emission from the model results,
   following the approach outlined in \citet{orl15}. Here we summarize 
   the main steps of this approach (see Sect. 2.3 in \citealt{orl15}
   for more details; see also \citealt{orl09,dra10}). 
   
   First, we rotated the system 10º about the $z$-axis, 
   60º about the $y$-axis, and 28º about 
   the $x$-axis to fit the orientation of the toroidal structure with 
   respect to the line of sight (LoS) found from the analysis of 
   observational data \citep{tro06,tro08}.
   For each cell of the 3D domain we derived the following:
   (a) the emission measure as $EM = n_{\mathrm{H}}^2 V$ 
   (where $n_{\mathrm{H}}$ is the hydrogen density and $V$ 
   is the volume of the emitting plasma; we assume fully ionized plasma);
   (b) the ionization age $\tau = n_\mathrm{ej} \Delta t$
   where $\Delta t$ is the time since the plasma in the domain 
   cell was shocked; and
   (c) the electron temperature $T_\mathrm{e}$, calculated from 
   the ion temperature, plasma density, and $\Delta t$, by 
   assuming Coulomb collisions and starting from an electron 
   temperature at the shock front $kT = 0.3$~keV, which is 
   assumed to be the same at any time as a result of
   instantaneous heating by lower hybrid waves 
   (\citealt{gha07}; see also \citealt[][for further details]{orl15}).
   In calculations of the electron heating and ionization timescale,
   the forward and reverse shocks are treated in the same way.
   From the values of emission measure, maximum ionization age,
   and electron temperature derived, we synthesized the X-ray 
   emission in the soft [$0.5-1.4$]~keV and hard [$1.4-5$]~keV
   bands using the non-equilibrium of ionization (NEI) emission model 
   VPSHOCK available in the XSPEC package along with the NEI 
   atomic data from ATOMDB 3.0 \citep{smi01}.

   We assumed a distance of 1.5~kpc to the source \citep{wel03},
   and an interstellar column density 
   $N_{\mathrm{H}}=7\cdot 10^{21}$~cm$^{-2}$ \citep{tro06}. 
   We assumed solar abundances for the ISM/CSM. As for the ejecta, 
   we synthesized the X-ray emission by assuming the following set of 
   abundances: O/O$_\odot$=6, Ne/Ne$_\odot$=7.5, Mg/Mg$_\odot$=7.5, 
   Si/Si$_\odot$=8, S/S$_\odot$=8, Fe/Fe$_\odot$=2.5, solar abundances 
   elsewhere. It is typically difficult to get reliable estimates of 
   the absolute abundances from the analysis of X-ray spectra 
   \citep[see][and references therein]{gre20}, while relative abundances 
   are much more robust. Our relative abundances are in agreement with 
   those derived by \cite{tro06,tro08} from the analysis of 
   \textit{XMM-Newton} observations (absolute abundances are a factor 
   $\sim5$ higher, see ``region 6'' in \citealt{tro06} and ``region IN'' 
   in \citealt{tro08}).
   We calculated the total X-ray emission in each cell and integrated along 
   the LoS, and in selected spatial regions for the synthetic spectra.
   We then folded the resulting emission through the instrumental response 
   of \textit{XMM-Newton}/EPIC MOS2 camera. 

   We also derived the distribution of emission measure vs.
   temperature, $EM(T)$, in the temperature range $[10^{4}-10^{8}]$~K. 
   The $EM(T)$ distribution is an important source of information
   on the plasma components with different temperature contributing
   to the emission and is very useful in order to compare model results 
   with observations. From the 3D spatial distributions of $T$ and $EM$, 
   we derive the $EM(T)$ for the computational domain as a whole 
   or for part of it: we consider the temperature range $[10^{4}-10^{8}]$~K 
   divided into 80 equidistant bins in log $T$; the total $EM$ in each 
   temperature bin is obtained by summing the emission measure 
   of all the fluid elements corresponding to the same temperature bin.

\section{Results}
\label{sec:results}

\subsection{Hydrodynamic evolution}
\label{sec:hd}

   \begin{figure*}
      \includegraphics*[width=\hsize]{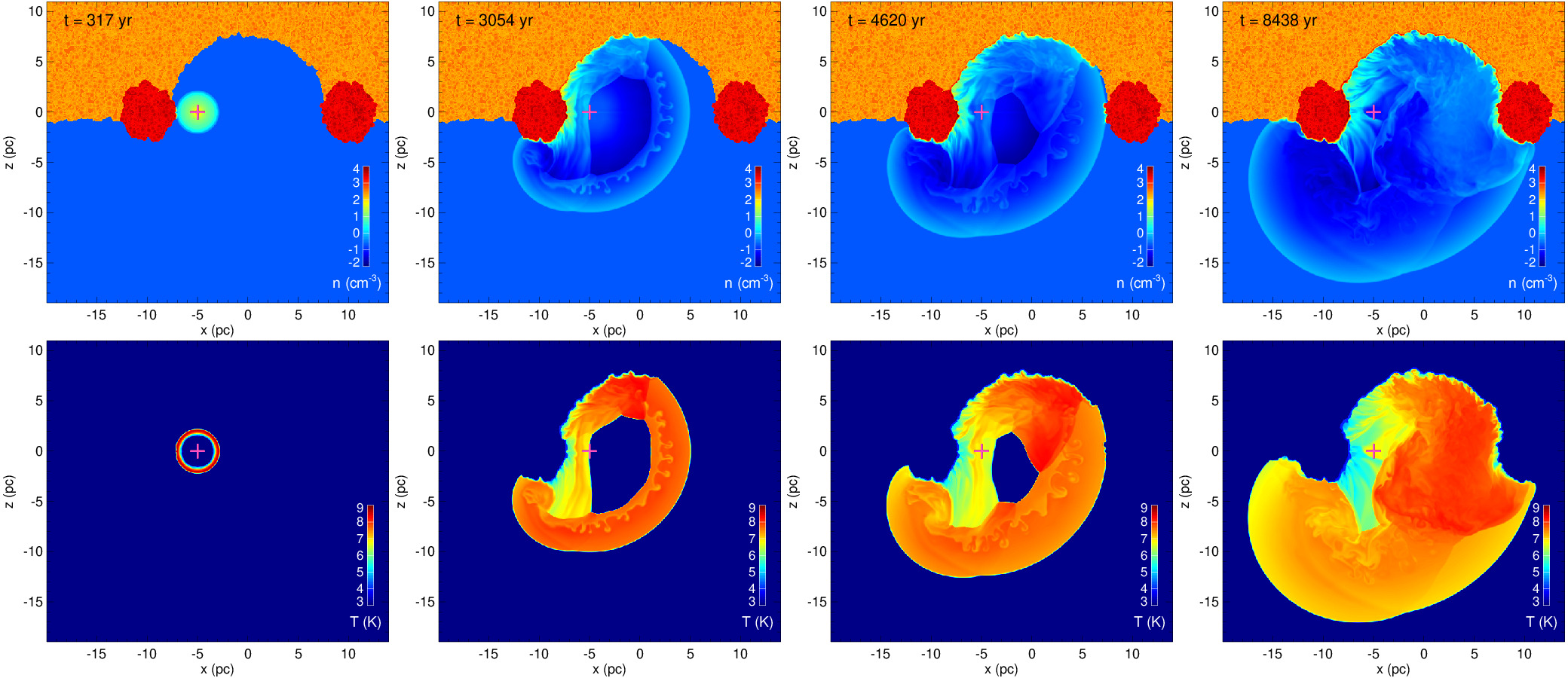}
      \includegraphics*[width=\hsize]{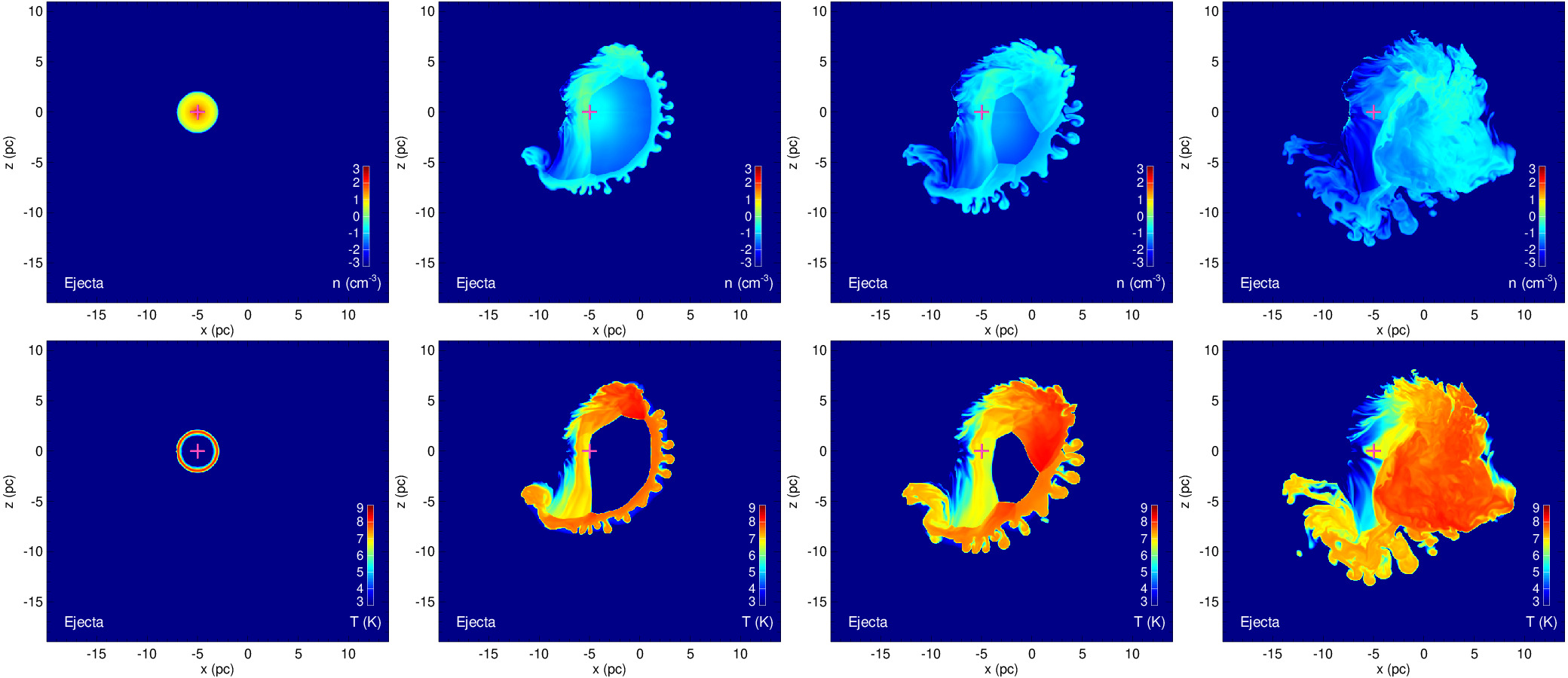}
      \caption{Density (first row; third row, only the ejecta) and ionic 
      temperature (second row; fourth row, only the ejecta) distributions 
      in logarithmic scale in the $(x,0,z)$ plane at different evolution 
      times (increasing from left to right).
      The magenta cross indicates the position of the explosion.
      See online Movie~1 for the complete temporal evolution.}
      \label{Fig:DensTemp}%
   \end{figure*}
   
   \begin{figure*}
      \includegraphics*[width=\hsize]{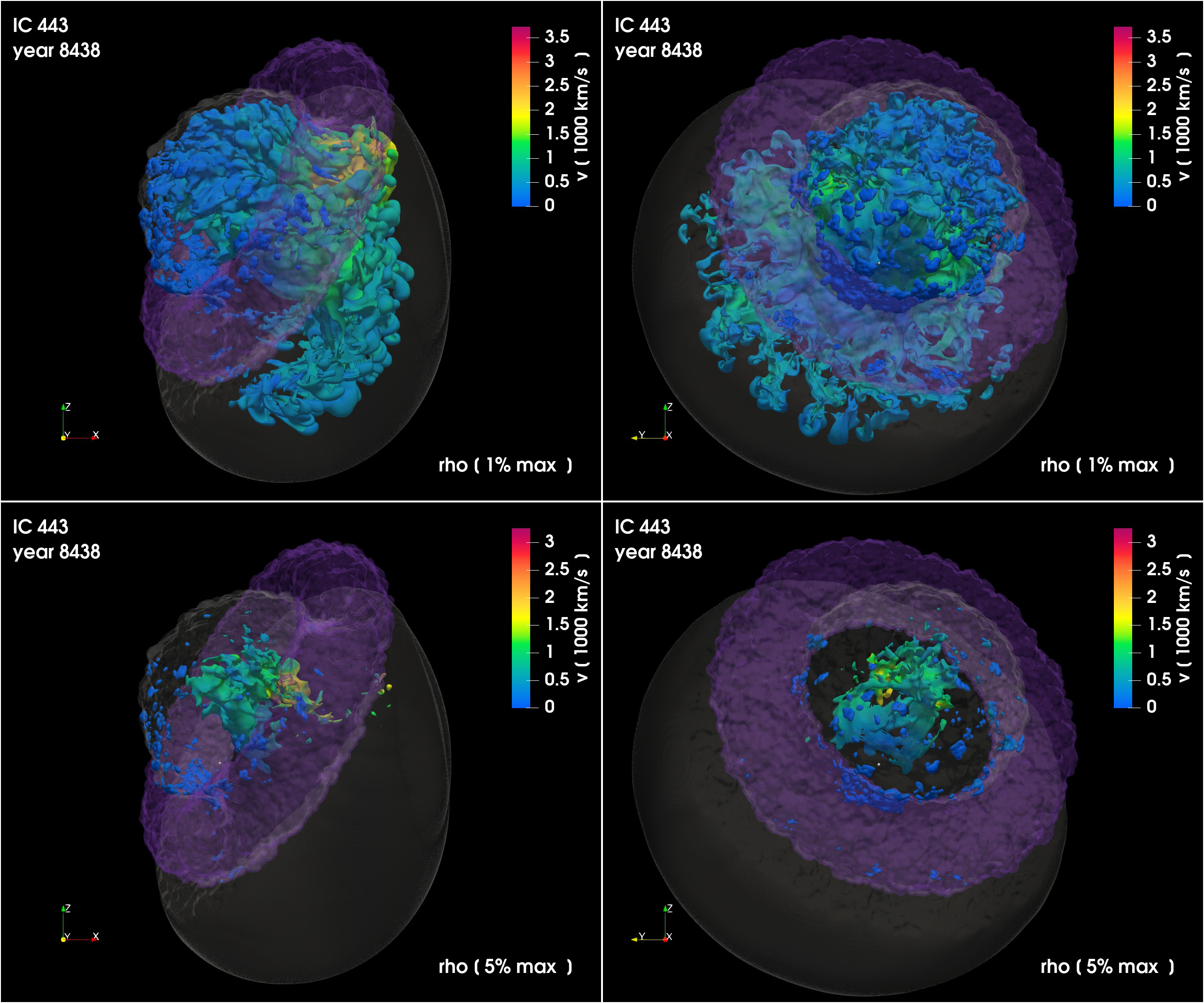}
      \caption{Isosurfaces of the distribution of density at $t\approx 8400$~yr 
      for the ejecta of the favourite model for SNR~IC~443 for different 
      viewing angles: along the LoS (left panels), and rotated by 90º about 
      the $z$-axis of the image (right panels).
      The opaque irregular isosurfaces correspond to a value of density 
      which is at 1\% (upper panels) and 5\% (lower panels) of the peak 
      density; their colors give the radial velocity in units of 
      1000~km~s$^{-1}$ on the isosurface.
      The semi-transparent surface marks the position of the forward shock;
      the toroidal semi-transparent structure in purple represents the 
      molecular cloud.
      See online Movie~2 and Movie~3 for an animation of these data;
      a navigable 3D graphic is available at \url{https://skfb.ly/6W9oM}.}
      \label{Fig:Ejecta}
   \end{figure*}
   
   \begin{figure}
      \includegraphics*[width=\hsize]{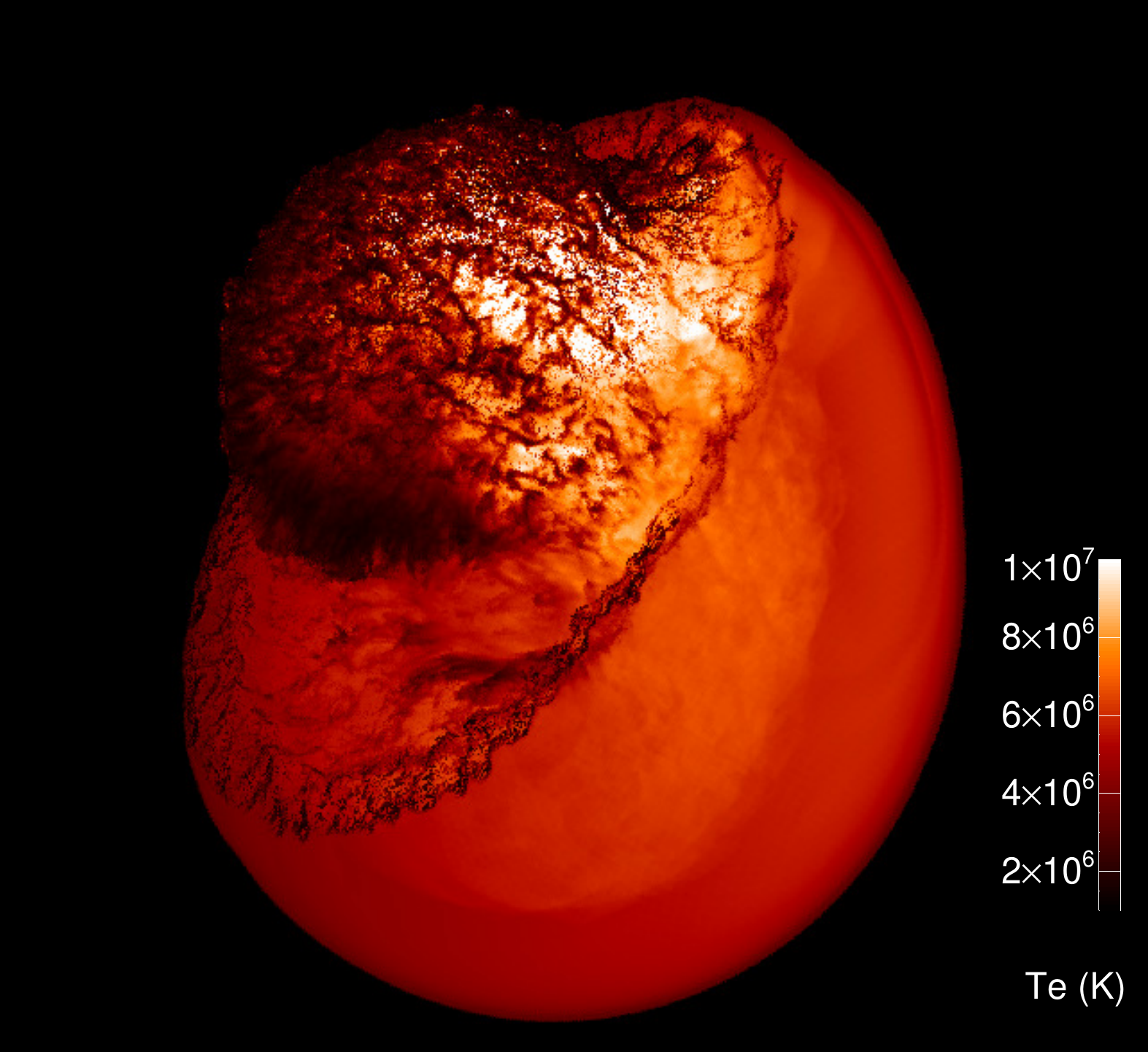}
      \caption{Map of the density-weighted average electronic temperature 
      integrated along the LoS at $\approx 8400$~yr.}
      \label{Fig:Tew}
   \end{figure}
   
    \begin{figure}
      \resizebox{\hsize}{!}{\includegraphics*{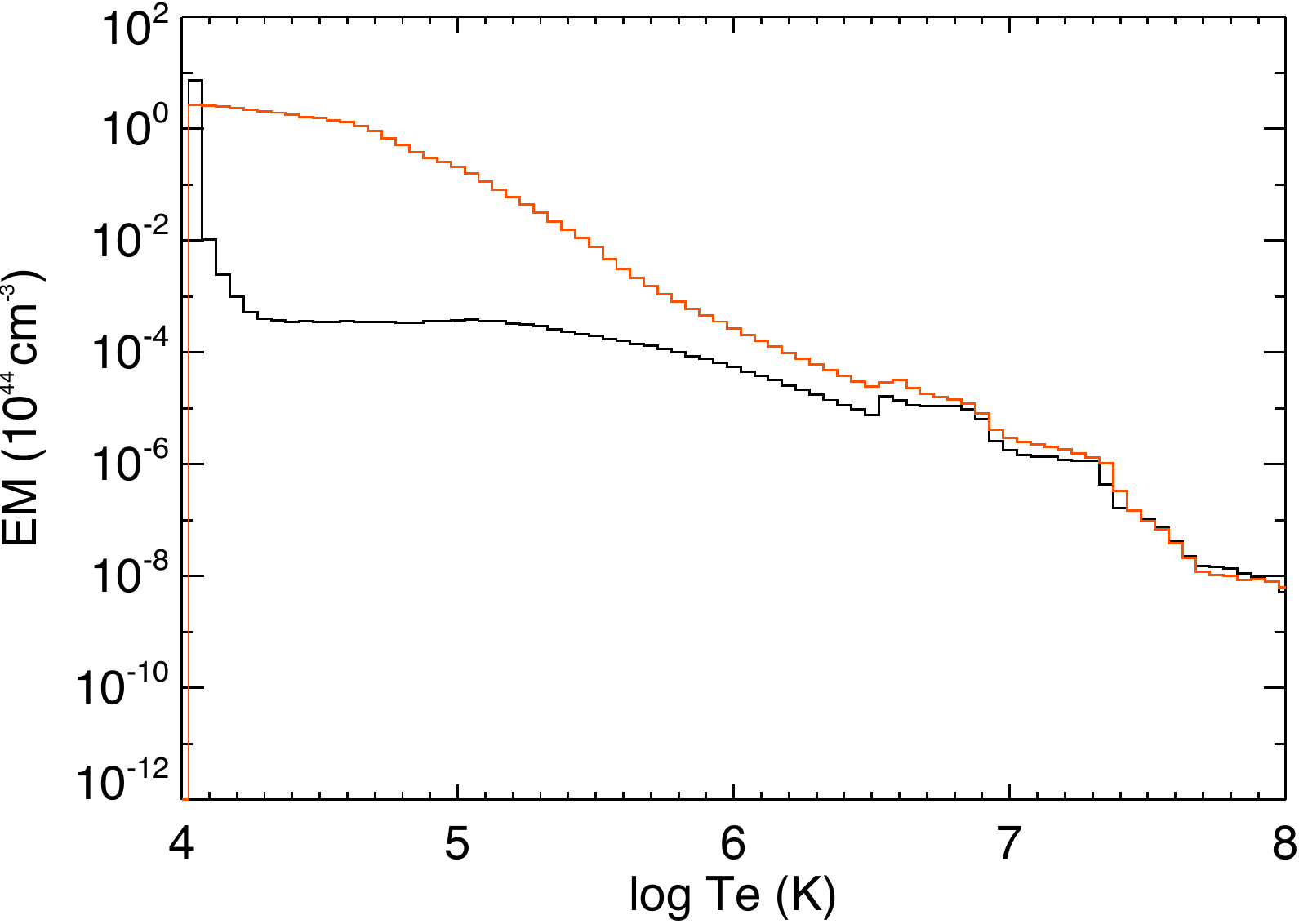}}
      \caption{Distribution of emission measure, $EM$, as a function of 
      the electronic temperature, T$_\mathrm{e}$, at $\approx 8400$~yr. 
      We compare our favourite model (black line) with the same model
      without radiative cooling (red line).}
      \label{Fig:EmTemp}
   \end{figure} 
   
   We followed the evolution of the SNR for approximately 10000~yr, 
   starting $\approx 100$~yr after the explosion. 
   We described the initial blast wave and the environment by performing a 
   wide exploration of the parameter space, including the mass of 
   the ejecta, the energy and origin of the explosion, and the density 
   of the clouds (see Table~\ref{parameters}). 
   Here we present in detail the results for our best model, 
   namely the case that best reproduces the observations of SNR~IC~443, 
   and discuss the exploration of the parameter space.
   
   The evolution of our favourite model is shown in Fig.~\ref{Fig:DensTemp}, 
   which reports the density (first row; third row, only the ejecta) and 
   temperature (second row; fourth row, only the ejecta) distributions in 
   the $[x,z]$ plane, in logarithmic scale, at different epochs (increasing 
   from left to right). The complete temporal evolution is available as 
   online movie (Movie~1).
   The ejecta, described with a spherically symmetric distribution at the 
   age of $\sim 100$~yr, initially propagate through a homogeneous medium in 
   all the directions powered by their high kinetic energy. 
   About $\sim 300$~yr after the SN event, the forward shock starts to 
   interact with the closer part of the toroidal molecular cloud (red 
   clumpy structure in the left top panel in Fig.~\ref{Fig:DensTemp}). 
   The interaction determines a strong slowdown of the forward shock hitting 
   the cloud and, consequently, a strengthening of the reverse shock 
   traveling through the ejecta. Meanwhile, the portion of the remnant not 
   interacting with the cloud continues to expand through the uniform 
   intercloud medium. As a result the initially spherically symmetric 
   remnant progressively becomes asymmetric in the subsequent evolution 
   (see second column in Fig.~\ref{Fig:DensTemp}). The asymmetry is even 
   enhanced when the blast wave starts to interact also with the atomic 
   cloud to the north (orange clumpy structure in the top panels in 
   Fig.~\ref{Fig:DensTemp}). The strong reverse shock powered by the 
   interaction with the molecular and atomic clouds (on the left side of 
   the domain) rapidly expands through the ejecta and crosses the position 
   of the neutron star immediately after the explosion (as also suggested 
   by \citealt{gre18}) at an age of $\approx 3000$~years (see the magenta 
   cross in the second column in Fig.~\ref{Fig:DensTemp}). 
   Given the large asymmetry caused by the interaction with the clouds on 
   the left side, the reverse shock does not refocus in the center of the 
   explosion, but proceeds expanding to the right (see third column in 
   Fig.~\ref{Fig:DensTemp}). At the same time, the forward shock continues 
   to expand, reaching the farthest part of the toroidal cloud at 
   $t\approx 4600$~yr (see third column in Fig.~\ref{Fig:DensTemp}). 
   Conversely, the portion of the blast wave traveling southward is free 
   to expand through the uniform intercloud medium. Most of the ejecta remain 
   confined by the dense clouds in the northern part of the domain, to the 
   right with respect to the center of the explosion (see the densest part 
   of the ejecta in the third row in Fig.~\ref{Fig:DensTemp}), while a 
   fraction of them expand freely to the south (see third and fourth rows 
   in Fig.~\ref{Fig:DensTemp}).
   
   The mass of the ejecta determines the density at the blast wave, and 
   together with the velocity regulates the energy of the explosion. 
   The mass of the ejecta in our favourite model is $\sim 7\,M_\odot$ and 
   the energy is $\sim 10^{51}$ erg. The higher is the mass of the ejecta 
   the higher is the density and the contribution of the ejecta to the 
   X-ray emission (see Sec.~\ref{sec:xray}). 
   Moreover, if we consider models with the same energy, the higher is
   the ejecta mass the longer takes the blast wave to expand and 
   the reverse shock to refocus; e.g., the model with ejecta 
   mass 10~$M_\odot$ takes $\sim 1000$~yr longer than the model with 
   4~$M_\odot$ to arrive to an analogous final evolution state. 
   When we consider models with higher energy instead (and same ejecta mass), 
   the blast wave expands faster due to the higher velocity of the ejecta 
   and the stronger interaction with the clouds leads to a larger 
   contribution to the X-ray emission (see Sec.~\ref{sec:xray}). For
   instance, the model with an explosion energy of $2.5\times 10^{51}$~erg
   takes $\sim 3000$~yr shorter than the model with $1\times 10^{51}$~erg
   to arrive to an analogous final evolution state.
   In the best model we consider an offset of 5~pc from the center 
   of the cavity for the progenitor (see Sec.~\ref{sec:num}), 
   corresponding to the position inferred by \cite{gre18} for the nearby 
   PWN at the instant of the explosion. A more centered position for the 
   origin of the explosion (see Appendix~\ref{app:sim}) yields a quite 
   symmetric distribution of material in the SNR that does not agree at all 
   with the multi-wavelength observations \citep{rho01,tro06,tro08,lee12,su14,gre18}.
   
   In Figure~\ref{Fig:Ejecta}, we show a 3D volume rendering of the ejecta
   in our favourite model for IC~443 at $t\approx 8400$~yr, from two 
   different points of view: on the left, the assumed LoS (see 
   Sect.~\ref{sec:synthesis}, and \citealt{tro06,tro08}); on the right, 
   same representation rotated by 90º about the $z$-axis. The ejecta 
   distribution considers cells with density at more than 1\% (upper panels) 
   and 5\% (lower panels) of the peak density. For a full-view distributions, 
   see the online animations (Movie~2 and Movie~3).
   We also compare the modeled distribution of the ejecta and the 
   blast wave with the observed morphology of the remnant in the optical 
   band in the online Movie~4 (the wide field optical image was provided 
   by Bob Franke - Focal Pointe Observatory) and in a navigable 3D graphic 
   (available at \url{https://skfb.ly/6X6BV}).
   The ejecta are confined by the toroidal molecular (purple semi-transparent 
   structure) and atomic clouds in the NE area, while the forward 
   shock (transparent surface) travels freely in the opposite direction.
   The range of values explored for the atomic cloud is $[10,10^3]$~cm$^{-3}$, 
   according to the values discussed in the literature (e.g. \citealt{rho01}); 
   for models with $n_\mathrm{a} < 300$~cm$^{-3}$ the forward shock sweeps 
   up part of the material of the cloud and produces higher X-ray emission 
   (see Sec.~\ref{sec:xray} for more details).
   In general, models with cloud densities $\gtrsim 300$~cm$^{-3}$ show 
   similar distributions of ejecta and X-ray emission (see Sec.~\ref{sec:xray}); 
   due to the high density the cloud acts as a wall where the forward shock 
   reflects. Thus, we selected for our best model the minimum density 
   producing results consistent with the observations, namely 
   $\sim 300$~cm$^{-3}$. We interpreted this as a lower limit to the density 
   of the atomic cloud, thus constraining better the values of density 
   quoted in the literature (namely $10<n_\mathrm{a}<10^3$~cm$^{-3}$; 
   e.g. \citealt{rho01}).
   As for the molecular cloud, we explored values in the range 
   $[10^3,10^4]$~cm$^{-3}$; for all of them, we found that neither the 
   ejecta nor the X-ray distributions show relevant differences. 
   So we decided to consider a density of the molecular cloud which is an 
   order of magnitude higher than that of the atomic cloud, namely 
   $\sim 3000$~cm$^{-3}$.
   
   We observe a very irregular and asymmetric distribution of the ejecta
   (see Fig.~\ref{Fig:Ejecta}) mainly due to the highly inhomogeneous medium
   where the SNR evolve and to the off-centered position selected for the
   origin of the explosion; the denser material is in the central part 
   of the remnant to the right of the center of the explosion inside the 
   cavity formed by the clouds (see 3rd and 4th rows in 
   Fig.~\ref{Fig:DensTemp}). This is due to the fact that the forward 
   shock reflects first on the NE clouds, is driven rapidly in the opposite 
   direction, shocking the ejecta expanding to the right, and, finally, 
   interacts with the reverse shock developed by interaction of the blast 
   wave with the NW clouds (note the color scale in Fig.~\ref{Fig:Ejecta} 
   and Movie~2, indicating the velocity distribution of the ejecta; see 
   also Fig~\ref{Fig:DensTemp}). As a result, the reverse shock converges 
   on a region located to the right of the center of explosion where the 
   ejecta are compressed, reaching the highest values of density.
   
   Since the SN explosion occurs close to the SE side of the molecular cloud, 
   ejecta are heated several times by multiple shocks reflected on the 
   surrounding clouds in the very early stages of the remnant evolution, 
   while they are still dense enough to approach rapidly the collisional 
   ionization equilibrium. After this early heating, which occurs close to 
   the explosion site, ejecta expand freely in the NW direction (see 
   Fig.~\ref{Fig:DensTemp} and Movie~1) and cool down adiabatically, 
   possibly leading to overionized plasma. This supports the scenario 
   proposed by \citet{gre18} to explain the presence of overionized plasma 
   and the component in strong  non-equilibrium of ionization (NEI) found 
   in the jet-like feature of IC~443.
   
   In Figure~\ref{Fig:Tew} we plot the density-weighted average electronic
   temperature along the LoS (see Sec.~\ref{sec:synthesis}) for our favourite 
   model at $t\approx 8400$~yr. 
   The temperature in the northern area close to the atomic cloud is 
   $\approx 2\times 10^6$~K, while in the SW region is higher, and reach 
   values of $\approx 10^7$~K in the central part of the remnant (close to 
   the molecular cloud). Thus, the figure evidences a rather complex 
   temperature structure of the plasma which is expected to be reflected in 
   the X-ray emission arising from the remnant, and a morphology which shows 
   a striking resemblance with that of IC~443.
   
   Figure~\ref{Fig:EmTemp} shows the distribution of emission measure
   as a function of the electronic temperature, $EM(T_\mathrm{e})$, 
   for our favourite model (black line), and the ideal case without 
   considering the effect of radiative losses from optically thin plasma 
   (red line). The two models show completely different distributions for
   T$_\mathrm{e} < 3 \times 10^6$~K (indicating an important role played 
   by the radiative cooling), while they appear to be very similar for 
   higher temperatures. Thus the density-weighted temperature and the 
   emission maps in the X-rays are similar for both the ideal and the best 
   case; the differences in the $EM(T_\mathrm{e})$ distributions for 
   T$_\mathrm{e} < 3 \times 10^6$~K are reflected mainly in the contribution 
   to the X-ray emission coming from the plasma interacting with the clouds, 
   which is higher in the ideal case.
   For a detailed description of the X-ray emission see Sec.~\ref{sec:xray}.

\subsection{X-ray emission}
\label{sec:xray}
   
   \begin{figure*}
      \includegraphics*[width=\hsize]{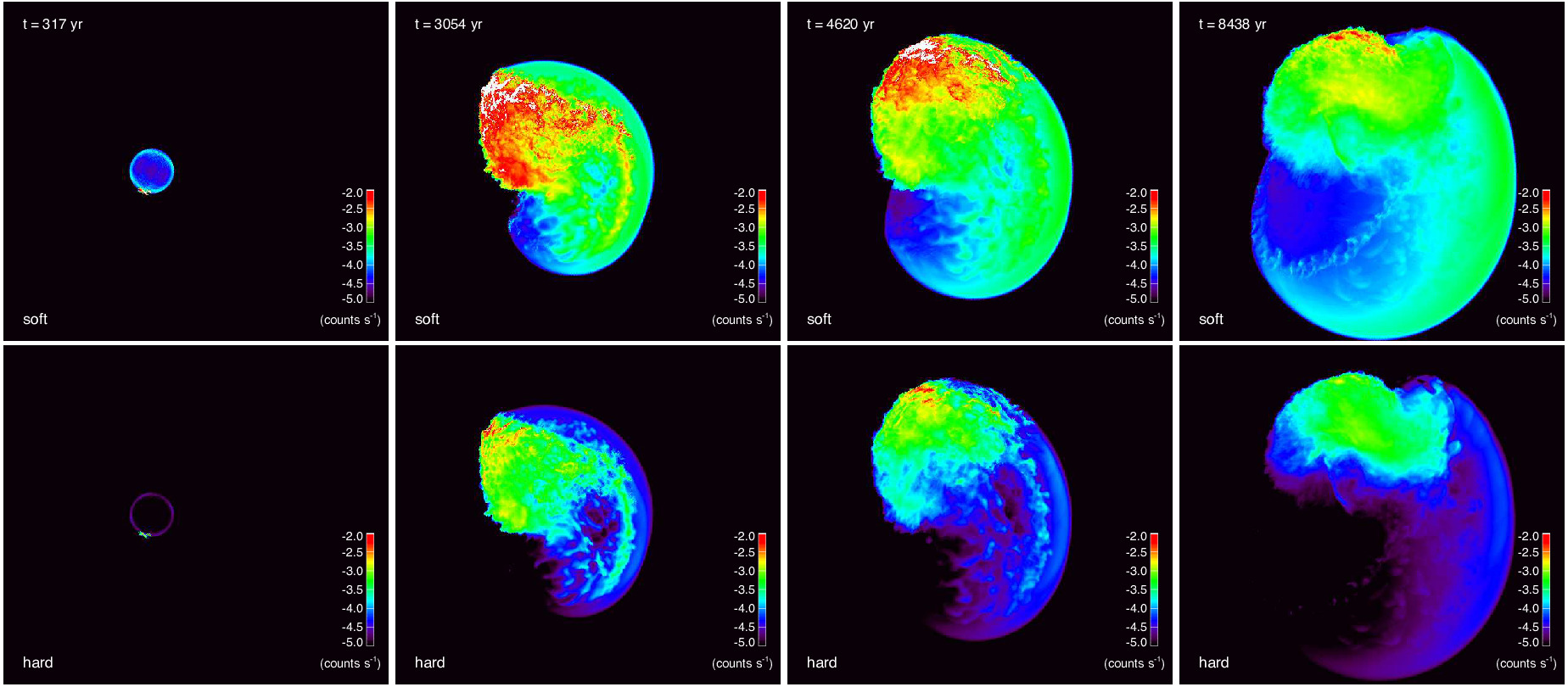}
      \caption{Synthetic X-ray count rate maps in the [$0.5-1.4$]~keV band 
      (upper panels) and [$1.4-5$]~keV band (bottom panels) in logarithmic 
      scale, at different evolution times (increasing from left to right).
      See online Movie~5 for the complete temporal evolution.}
      \label{Fig:Xray}%
   \end{figure*}
   
   \begin{figure*}
      \includegraphics*[width=\hsize]{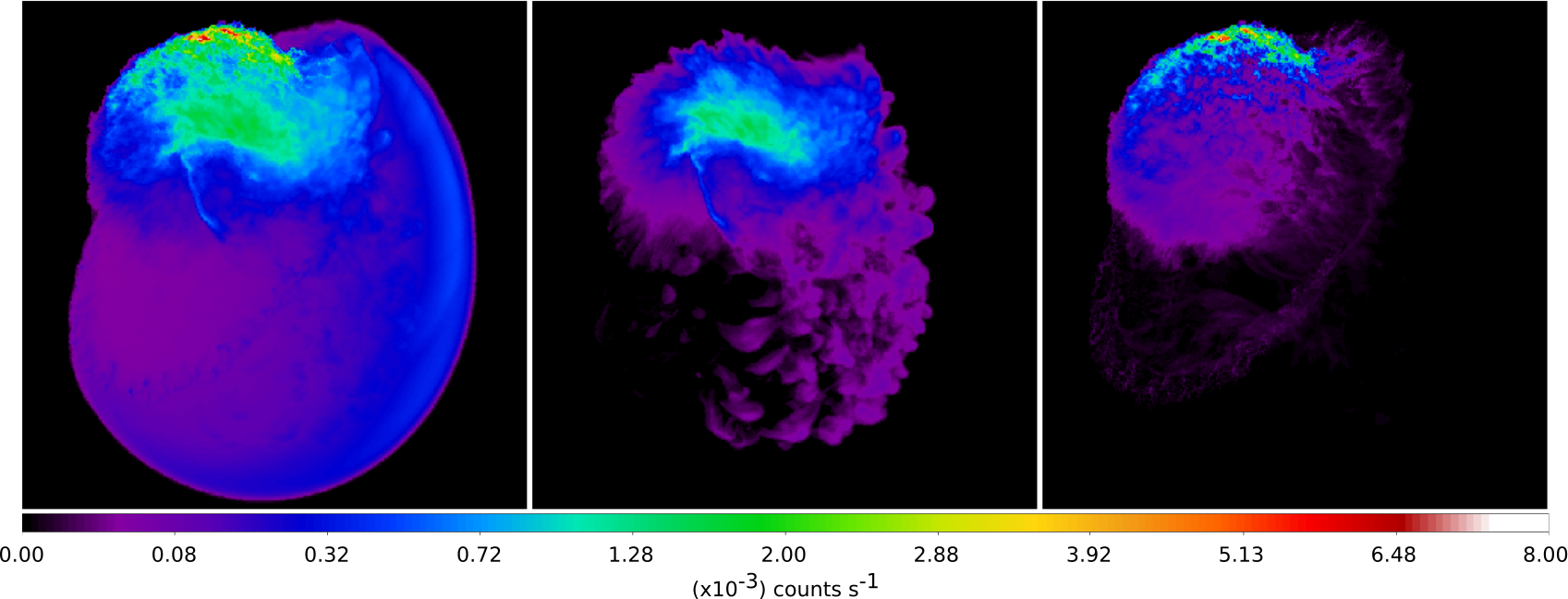}
      \caption{Synthetic X-ray count rate maps in the
      [$0.5-1.4$]~keV band (in square root scale) at $t\approx 8400$~yr, 
      derived from the best model as explained in Sec.~\ref{sec:synthesis} 
      for the different components of the plasma: the whole distribution 
      (left panel), only the ejecta (middle panel), and the atomic and 
      molecular clouds (right panel).}
      \label{Fig:XraySoft}
   \end{figure*}
   
   \begin{figure*}
      \includegraphics*[width=\hsize]{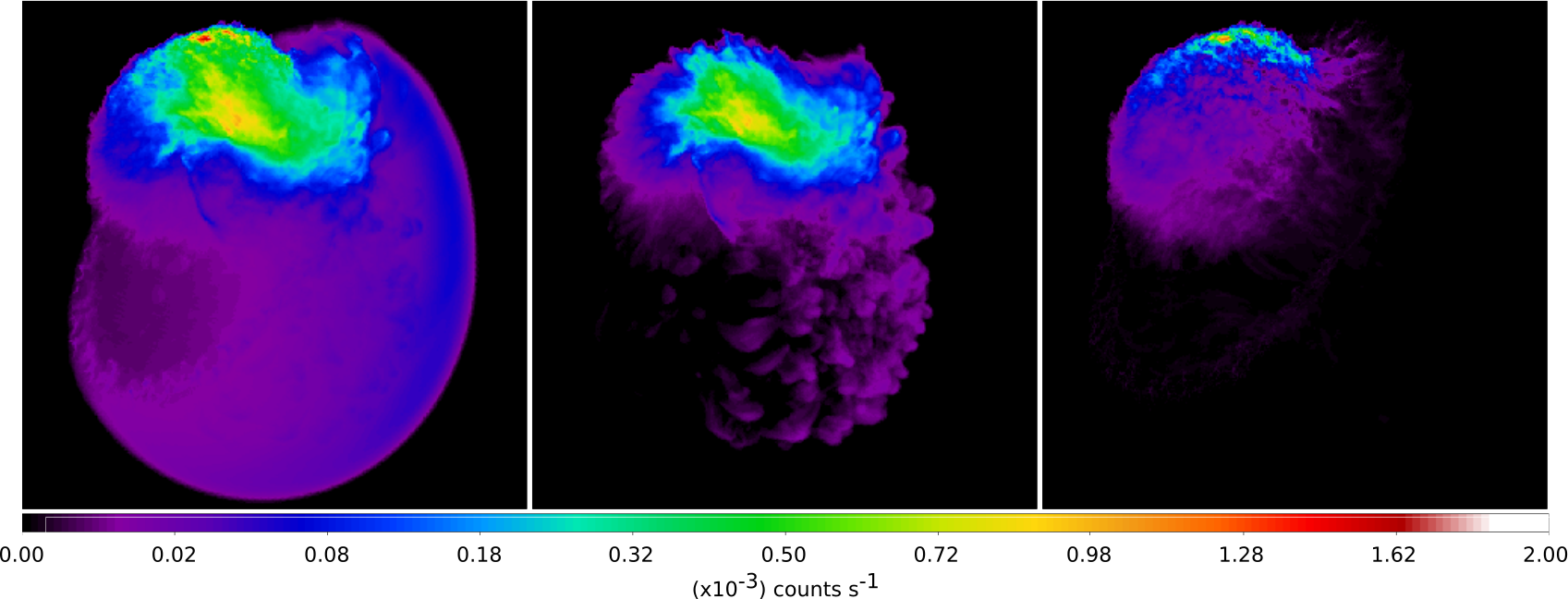}
      \caption{Synthetic X-ray count rate maps in the
      [$1.4-5$]~keV band (in square root scale) at $t\approx 8400$~yr, 
      derived from the best model as explained in Sec.~\ref{sec:synthesis} 
      for the different components of the plasma: the whole distribution 
      (left panel), only the ejecta (middle panel), and the atomic and 
      molecular clouds (right panel).}
      \label{Fig:XrayHard}
   \end{figure*}
   
   \begin{figure*}
      \includegraphics*[width=\hsize]{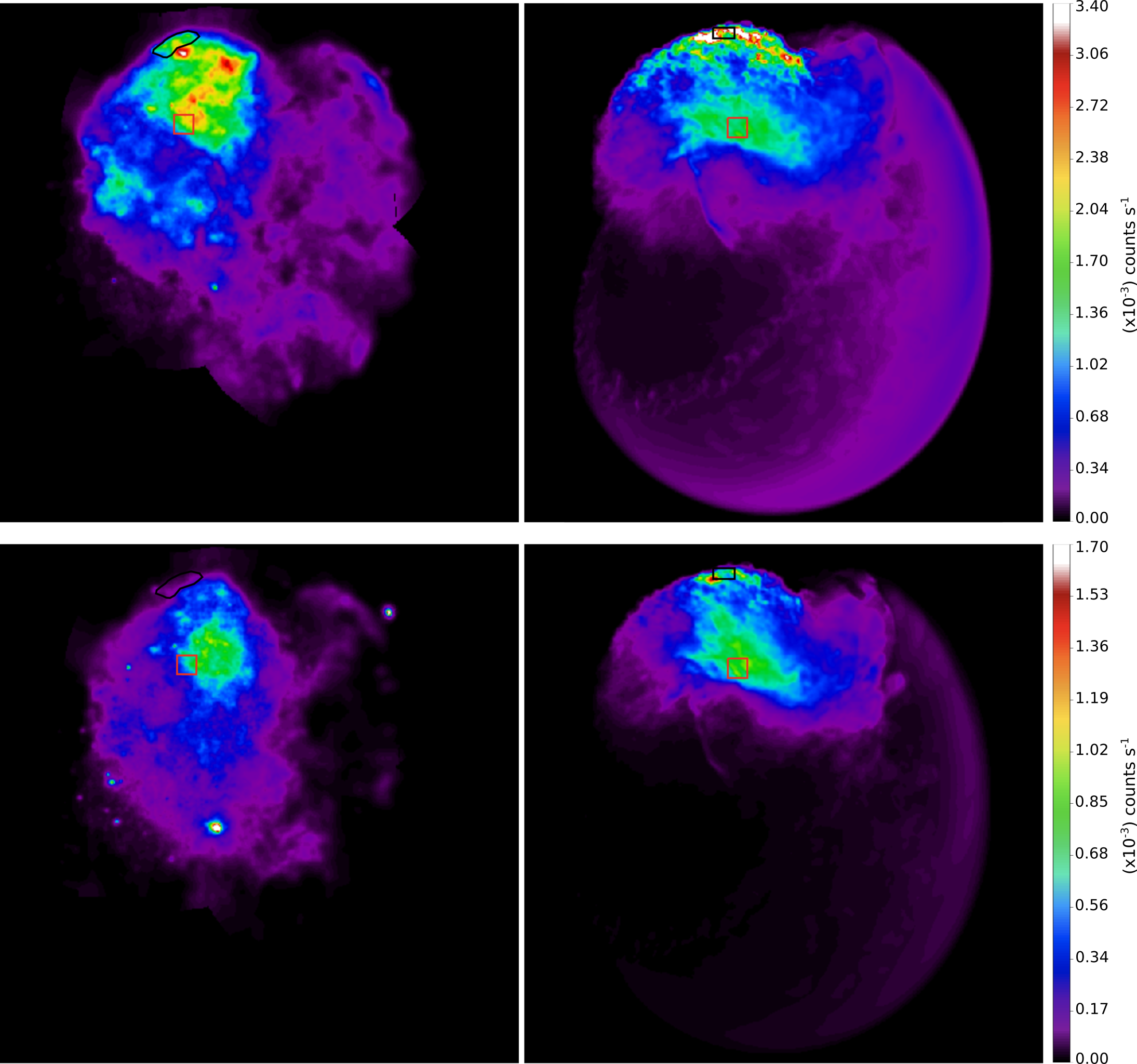}
      \caption{Smoothed X-ray count rate maps in the [$0.5-1.4$]~keV 
      (upper panels) and [$1.4-5$]~keV (lower panels) bands with a pixel 
      size of 11\arcsec. 
      Left panels: 2010 observations resampled as explained in the 
      Appendix~\ref{app:obs}. 
      Right panels: synthetic images derived from the model as explained 
      in Sec.~\ref{sec:synthesis}.
      The black and red shapes indicate the regions selected for the 
      spectra shown in Fig.~\ref{Fig:Spec}.}
      \label{Fig:XrayObsMod}
   \end{figure*}
   
   \begin{figure*}
      \includegraphics*[width=0.5\textwidth]{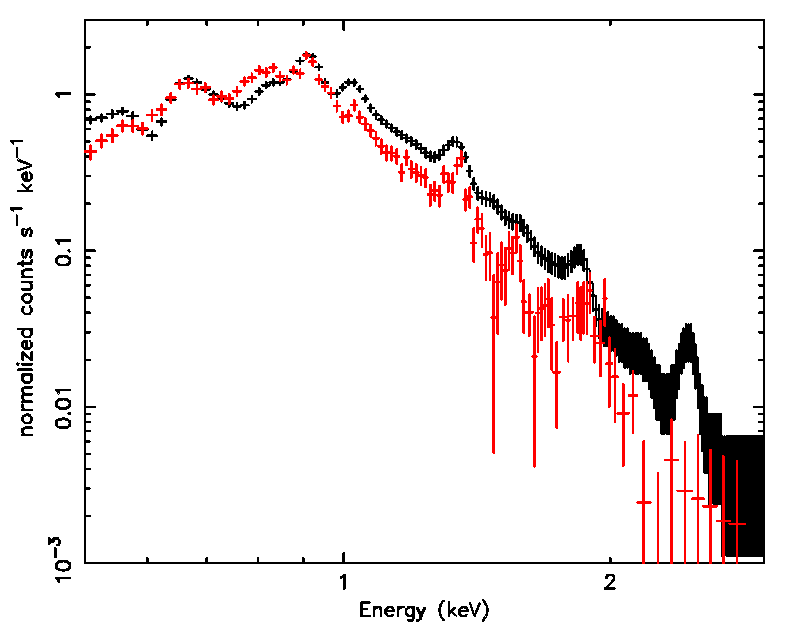}
      \includegraphics*[width=0.5\textwidth]{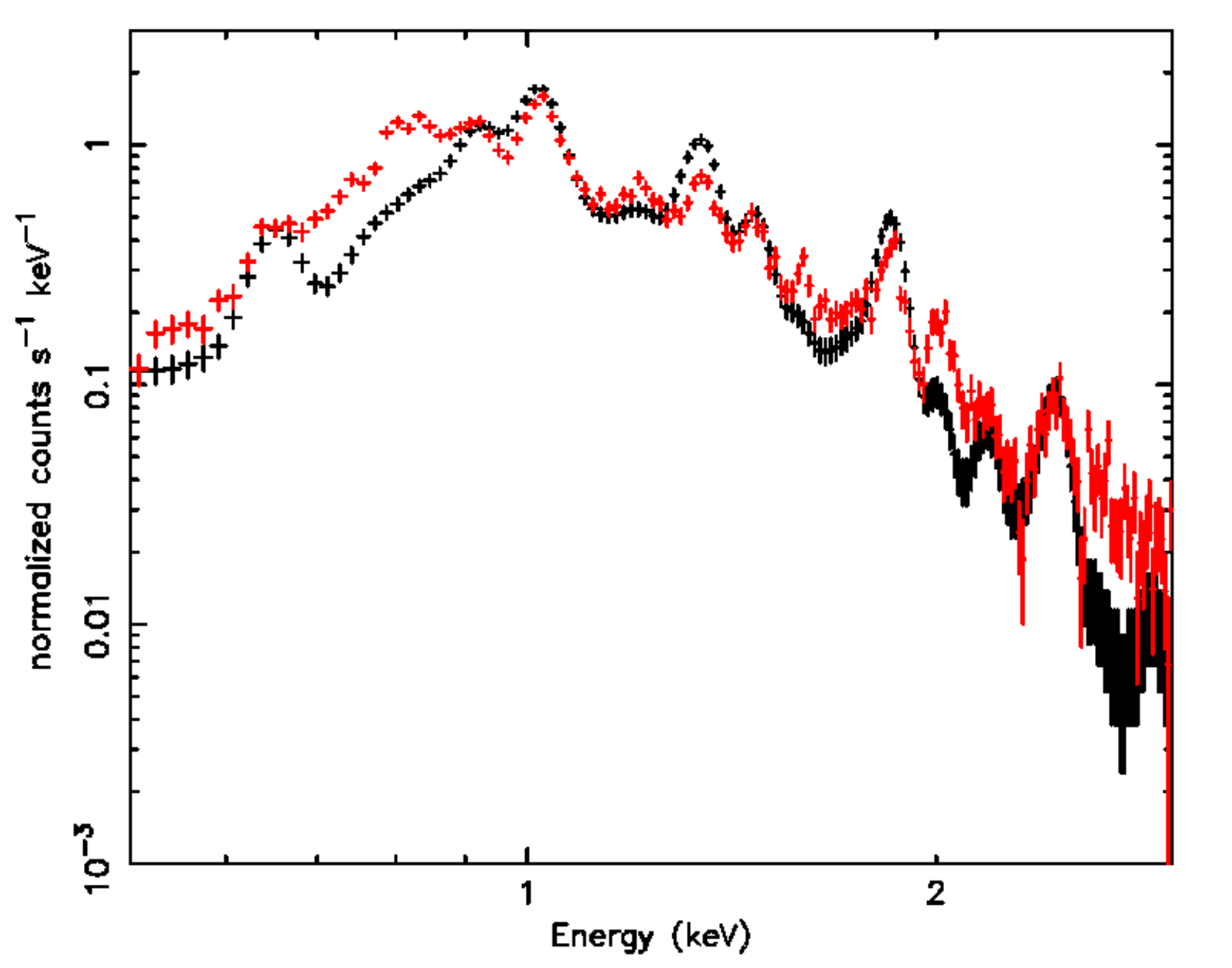}
      \caption{\emph{XMM-Newton} EPIC MOS spectra extracted from the northern 
      (left panel) and central (right panel) regions represented in 
      Fig.~\ref{Fig:XrayObsMod}. 
      In black, the synthetic spectra derived from the model; in red, 
      the spectra extracted from the observations.
      Error bars are at 1$\sigma$.}
      \label{Fig:Spec}
   \end{figure*}
   
   In this section we analyse the distribution of the X-ray emission
   in our best model, and compare the X-ray emission synthesized from the 
   model (see Sec.~\ref{sec:synthesis} for a detailed description) with 
   \textit{XMM-Newton} observations of SNR~IC~443 (see Appendix~\ref{app:obs} 
   and \citealt{gre18} for more details). 
   
   In Figure~\ref{Fig:Xray} we present X-ray count rate maps synthesized
   from the model at different evolution times (increasing from left to 
   right, corresponding to the evolution times shown in 
   Fig.~\ref{Fig:DensTemp}), considering the LoS as in left panels in 
   Fig.~\ref{Fig:Ejecta} and integrating the emission along the LoS (see 
   Sec.~\ref{sec:synthesis}).  
   The complete temporal evolution is available as online movie (Movie~5).
   The upper and lower panels at each epoch show the soft ([$0.5-1.4$]~keV),
   and hard ([$1.4-5$]~keV) X-ray emission, respectively.
   When the expanding ejecta reach the molecular cloud at 
   $t\approx 300$~yr since the SN event, the interaction of the forward shock 
   with the dense cloud produces bright X-ray emission in both soft and hard 
   bands (see left panels in Fig.~\ref{Fig:Xray}). The impact on the 
   cloud produces a reflected shock that powers the reverse shock, heating 
   the ejecta material (see Fig.~\ref{Fig:DensTemp});
   indeed, the second column panels in Fig.~\ref{Fig:Xray} evidences 
   a great area with strong X-ray emission, especially in the soft band,
   due to the interaction of the ejecta with the dense cloud and the powered 
   reverse shock. The forward shock continues to interact with the atomic 
   cloud to the north while expanding and producing bright X-ray emission 
   that moves clock-wise from east to the north during the evolution (see 
   Fig.~\ref{Fig:Xray}, from left to right, and Movie~5).
   The free propagation of the forward shock to the SW heats the ambient 
   interclump material producing soft, and faint hard, X-ray emission
   (see third and fourth columns, in Fig.~\ref{Fig:Xray}).
   We find the brightest X-ray emission in the north area of the 
   remnant where the clouds keep the plasma confined.
   
   We followed the evolution and calculated the X-ray emission of the 
   different plasma components (namely ISM and ejecta) by following the 
   procedure explained in Sec.~\ref{sec:synthesis}. In 
   Figures~\ref{Fig:XraySoft}~and~\ref{Fig:XrayHard} we show synthetic
   X-ray count rate maps, in the soft and the hard bands respectively (note
   the different scales), for the best model at $t\approx 8400$~yr.
   Similarly to the density distributions shown in 
   Fig.~\ref{Fig:DensTemp}~and Fig.~\ref{Fig:Ejecta}, the X-ray emission maps
   show a very asymmetric shape due to the highly inhomogeneous medium 
   where the SNR evolve and to the off-centered position selected for the
   origin of the explosion; due to the morphology of the surrounding clouds,
   a more centered position in the cavity for the originating SN 
   derives in a more symmetric distribution (see Fig.~\ref{Fig:XrayC} and 
   Appendix~\ref{app:sim}) that does not agree with 
   the X-ray studies performed \citep{tro06,tro08,boc09,gre18}.
   The ejecta show a centrally peaked distribution (see 
   Fig.~\ref{Fig:XraySoft}~and Fig.~\ref{Fig:XrayHard}, middle panels) in 
   both soft and hard bands. From the exploration of the parameter space, 
   we found out that the higher is the initial mass of the ejecta the 
   brighter is its X-ray distribution. 
   
   The ejecta and the interstellar material are confined in the NE area 
   (see Fig.~\ref{Fig:Ejecta}) and interact with the clouds emitting in 
   X-rays, mainly in the soft band, from the limb close to the atomic cloud 
   (see left panels in Fig.~\ref{Fig:XraySoft}~and~Fig.~\ref{Fig:XrayHard}).
   We verified that by changing the cloud density, we induce variations of 
   the intensity of the X-ray emission in the northeastern bright limb. 
   In particular, a lower cloud density leads to a higher contribution of 
   X-ray emission from the interaction with the clouds. In fact, in lower 
   density ($n_\mathrm{a} < 300$~cm$^{-3}$) models, the forward shock impacts 
   into the cloud and drags part of the material; this interaction produces 
   higher X-ray emission than the one shown in right panels in
   Fig.~\ref{Fig:XraySoft}~and Fig.~\ref{Fig:XrayHard}.
   On the other hand, for $n_\mathrm{a} > 300$~cm$^{-3}$, and the range of 
   values explored here for the density of the molecular cloud, namely 
   $[10^3,10^4]$~cm$^{-3}$, the X-ray distributions don't show relevant 
   differences. In any case, models with $E > 1\times 10^{51}$~erg, that 
   is the energy of the best case, show higher X-ray emission in both the 
   soft and the hard bands for all the density values of the clouds explored 
   in this paper (see Table~\ref{parameters}), due to the higher velocity of 
   the blast wave that leads to a stronger shock while interacting with the 
   clouds and the ejecta.
   
   SNR~IC~443 has been classified in the category of MMSNRs which show 
   thermal peaked X-ray emission in the inner part and a shell morphology in 
   radio \citep{rho98}. The origin of this peculiar distribution, observed 
   in left panels of Fig.~\ref{Fig:XrayObsMod}, is not well understood.
   The X-ray emission synthesized from our favourite model has a centrally 
   peaked component coming mainly from the ejecta in both soft and hard bands 
   (see right panels in Fig.~\ref{Fig:XrayObsMod}).
   By carefully deriving the explosion site (associated with 
   CXOU J061705.3+222127) and the evolutionary stage ($\sim 8000$ yr) of 
   IC~443, we can naturally reproduce the broadband shape of the remnant. 
   Moreover, we can understand the physical origin of the two different 
   morphologies observed in soft ($0.5-1.4$ keV) and hard ($1.4-5$ keV) X-rays. 
   In particular, we can pinpoint the role of ejecta and ISM in shaping the 
   X-ray emission. While the interaction with the clouds produce mainly soft 
   X-rays, concentrated in the upper part of the remnant (at $t\sim 8000$ yr), 
   the hard X-ray emission is more centrally peaked (as in real data) and 
   includes a larger contribution from the shocked ejecta heated by a 
   reflected shock generated by the impact with the surrounding molecular ring. 
   This is in agreement with the spectral analysis performed by 
   \citet{tro06,tro08}, who associated the soft X-ray emission with shocked 
   ISM and the harder X-ray emission with ejecta. The peculiar V-shaped 
   northern border of the remnant and the faint, diffuse central and southern 
   emission are also reproduced by our simulation as a result of the topology 
   of the circumstellar environment.
   
   We point out that the synthetic count rate is pretty similar to that 
   actually observed (without the need of adding any ad-hoc renormalization), 
   as shown in Fig. \ref{Fig:XrayObsMod}. We just note that synthetic maps 
   show an excess of soft X-ray emission in the northern limb, due to the 
   strong component originating in the interaction of the SNR with the atomic 
   cloud (see Fig.~\ref{Fig:Xray} and Movie~5, and right panels in 
   Fig.~\ref{Fig:XraySoft}, Fig.~\ref{Fig:XrayHard} and 
   Fig.~\ref{Fig:XrayObsMod}).
   This excess is a result of the idealized morphology considered 
   for the atomic cloud. Indeed, we considered a spherical cap connected 
   with the toroidal molecular cloud that keeps the plasma confined and 
   causes several reflections that, in turn, heat the material trapped inside.
   The X-ray emission from SNR~IC~443 is higher in the soft band
   (see Fig.~\ref{Fig:XrayObsMod})\footnote{Note the different scales used
   for the soft (upper panels) and the hard (lower panels) X-ray bands.}
   due to the strong interaction of the remnant with the CSM/ISM.
   
   In order to get a deeper level of diagnostics, we compared synthetic and 
   actual spectra in the northern and central part of the remnant. 
   To this end, we synthesized X-ray spectra from the two regions shown in 
   right panels of Fig.~\ref{Fig:XrayObsMod}: the northern region of the 
   image being mainly composed by X-ray emission originating from the 
   SNR-cloud interaction (black), and the central region being dominated 
   by the ejecta component (red). We then analyzed actual spectra extracted 
   from corresponding regions in the real \emph{XMM-Newton} data, shown in 
   left panels of Fig.~\ref{Fig:XrayObsMod}, following the method outlined 
   in \cite{gre18} (see also Appendix~\ref{app:obs}).
   We selected regions with similar observed and synthetic total count rate, 
   in order to compare the shape and the main features of the spectra. 
   In Figure~\ref{Fig:Spec} we plot the spectra corresponding to the northern 
   (left panel) and the central (right panel) regions; in both cases, the 
   spectra derived from the best model (black) remarkably reproduce the main
   features of the spectra extracted from the observed data (red). 
   In particular, the equivalent widths of emission lines are much more 
   pronounced in the central region than at north. Moreover, the overall 
   slope of the continuum is steeper at the north than in the central region, 
   as in real spectra.
   Details of line emission differ between models and observations (see 
   Fig.~\ref{Fig:Spec}, right panel). This is because we do not include 
   tracers to follow the different chemical elements in the model, but just 
   impose uniform metallicity for all the ejecta in post-processing, 
   according to the relative abundances of the main elements derived by 
   \cite{tro06,tro08}.

\section{Summary and conclusions}
\label{sec:conclusions}

   In this work, we investigated the origin of the complex X-ray 
   emission observed in the mixed-morphology SNR~IC~443.
   For this purpose, we developed a 3D HD model describing 
   the interaction of the SNR with the environment, parametrized in 
   agreement with the results of the multiwavelength data analysis
   \citep{rho01,tro06,tro08,lee12,su14,gre18}. 
   We performed a wide exploration of the parameter space describing 
   the initial blast wave and the environment, including the mass of 
   the ejecta, the energy and origin of the explosion, and the density 
   of the clouds (see Table~\ref{parameters} for a summary of the 
   parameter space explored and the values best reproducing the observations).
   From the simulations, we synthesized the X-ray emission and compared it
   with \textit{XMM-Newton} observations (see Appendix~\ref{app:obs}).  

   Our model explains the complex X-ray morphology of SNR~IC~443 in a 
   natural way, being able to reproduce most of the features observed 
   and identifying the strong effect of the inhomogeneous ISM on the remnant. 
   The centrally-peaked X-ray morphology, is best reproduced when considering 
   the explosion site at the position of the PWN CXOU~J061705.3+222127 at the 
   time of the explosion (inferred taking into account the proper motion
   of the neutron star as estimated by \citealt{gre18}). 
   This fact, together with the jet-like structure detected by \cite{gre18}, 
   supports the association between the PWN and IC~443, strongly indicating 
   that the PWN belongs to IC~443 and that the collimated jet-like structure 
   has been produced by the exploding star. 
   
   Our model results in a very irregular and asymmetric distribution of 
   the ejecta, with a centrally-peaked X-ray emission, due to the highly 
   inhomogeneous medium where the SNR evolve and the off-centered position 
   selected for the origin of the explosion.
   The surrounding clouds form a cavity where multiple shocks reflect, 
   heating repeatedly the ejecta trapped inside. Before the expansion, the 
   ejecta are heated and ionized by the interaction with the reflected 
   shock due to the impact of the forward shock front with the SE cloud 
   (which is very close to the explosion site, indicated by the position
   of the PWN at that time). Afterwards, the ejecta may cool down via 
   adiabatic expansion to the NW, which could explain the overionized plasma 
   in the jet-like structure discovered by \cite{gre18}.
   
   In our model, the atomic and the molecular clouds confine the remnant to 
   the NE and SE, while it expands freely to the SW (see Fig.~\ref{Fig:Ejecta}
   and Movie~2 for a 3D view). From the exploration of the parameter space
   and the comparison of the X-ray emission observed with the X-ray 
   synthesized images, we find that the density of the atomic cloud should be
   $n_\mathrm{a} > 300$~cm$^{-3}$, thus constraining better the values of 
   density quoted in the literature (namely $10<n_\mathrm{a}<10^3$~cm$^{-3}$; 
   e.g. \citealt{rho01}). In models with $n_\mathrm{a} < 300$~cm$^{-3}$ the 
   forward shock sweeps up part of the material of the cloud and produces 
   higher X-ray emission. 
   For the molecular cloud, instead, neither of the values explored, namely 
   $[10^3,10^4]$~cm$^{-3}$, show significant differences in the X-ray emission.
   In our best model, the mass of the ejecta and the energy of 
   the explosion are $\sim 7\,M_\odot$ and $\sim 1\times 10^{51}$ erg, 
   respectively, indicating that the parent SN was characterized by a low 
   explosion energy. The higher is the mass of the ejecta and the energy of 
   the model, the brighter is the X-ray emission synthesized.
   From the evolution of the best model, we found that the peak of the 
   X-ray emission move with time clockwise from east to the north and the 
   time that best reproduces the X-ray observations of the remnant is 
   $\approx 8400$~yr. Considering all the cases explored we find that the 
   age of SNR~IC~443 is $\sim 8000$~yr, indicating that IC~443 could be 
   much younger than previous studies estimated \citep[e.g.][]{che99,byk08}.
   
   The X-ray emission from SNR~IC~443 is mostly dominated by the soft 
   band, with a centrally-peaked distribution from the ejecta, and a
   bright X-ray component from the northern limb close to the atomic 
   cloud (see upper panels in Fig.~\ref{Fig:XrayObsMod}). The X-ray emission 
   in the [$1.4-5$] keV band is dominated by the ejecta component and 
   presents a more pronounced centrally peaked morphology, as in actual data.
   The synthetic spectra derived from the model reproduce well the 
   main features of the spectra extracted from the observed data.
   
   It is also worth to mention that our simulations do not treat 
   molecular and atomic dissociation and ionization in the dense clouds 
   because these processes are negligible in the energy balance of the model. 
   Considering a composition of 90\% H and 10\% He for the clouds, we 
   estimated that the total energy necessary to dissociate and ionize all 
   the shocked atoms and molecules of the clouds at the age of $\sim 8400$~years is $\sim 10^{48}$~erg. This value is one order of magnitude lower than the total internal energy 
   of the shocked cloud material at the same epoch (which is $\sim 10^{49}$~erg).
   Thus, we do not expect that the inclusion of these effects could change 
   the main conclusions of this work. We also note that the internal structure of the remnant is mainly determined by the reverse shock that heats the ejecta, leading to the 
   centrally-peaked morphology that emerges in the X-rays (see Fig.~\ref{Fig:XrayObsMod}). Eventually, the main effect of dissociation and ionization of H and He would be to slightly reduce the temperature of the shocked clouds material, thus slightly lowering the contribution to the X-ray flux from the clouds in the [$0.5-1.4$]~keV soft band (see right panel in Fig.~\ref{Fig:XraySoft}). This reduction would improve even more the agreement between the X-ray emission maps synthesized from our model and those observed (see Fig.~\ref{Fig:XrayObsMod}).

   The radiative losses included in the model cool down the plasma, 
   although in every case the synthesized X-ray emission produced from 
   the interaction with the atomic cloud is brighter than that observed 
   in X-ray data. 
   The excess of X-ray emission produced in the interaction of the SNR
   with the clouds in our model is a result of the morphology considered
   for the atomic cloud. Indeed, we considered a spherical cap, connected 
   with the toroidal molecular cloud, that keeps the plasma confined and 
   produces several reflections that heat the material trapped inside.
   The morphology described here is an idealization of the environment
   that we expect in SNR~IC~443; a small fraction of the material could 
   escape from the bubble along the LoS without leaving clear evidence
   in the observations.
   
   It is worth to note that we neglected the effects of the thermal 
   conduction in our simulations, although it may play a role in 
   modifying the temperature structure of the shocked cloud and ejecta. 
   In fact, a crucial effect that should be taken into account when 
   including the thermal conduction is the ambient magnetic field which 
   makes the thermal conduction highly anisotropic. However we 
   do not have any hints on the configuration and strength of the pre-SN 
   interstellar magnetic field in IC~443. Any exploration of different 
   field configurations and strengths in the simulations would be hopeless
   without any constraints from observations. Nevertheless, even if we would 
   have a satisfactory knowledge of the field configuration, we think that 
   the effects of thermal conduction may be negligible. For example, in the 
   case of an ordered magnetic field, the thermal conduction can be highly 
   suppressed in the direction perpendicular to the magnetic field 
   (e.g. \citealt{orl08}) and this could be the case if the blast wave 
   squeeze the interstellar magnetic field along the clouds border 
   (e.g. \citealt{orl19}). In this case, the results would be similar 
   to those presented in the paper. In the case of a randomly oriented 
   magnetic field (possibly due to local turbulence) the thermal 
   conduction can be reduced by up to a factor of 5 \citep{nar01}, 
   thus making less relevant the effects of the thermal conduction. 
   Considering this reduction factor and the spatial resolution of 
   the simulations as the lowest temperature length-scale, we estimated the characteristic 
   time-scale for the thermal conduction in our model following the 
   Eq.~10 in \cite{orl05}. We obtained a conduction time-scale of 
   $\sim 5\times 10^5$ years, which is much larger than the evolution time 
   of our model ($\sim 9000$ years). Thus, rather than prescribing an 
   arbitrary magnetic field, for the purposes of the paper, we preferred 
   to neglect the effects of thermal conduction in our simulations. 
   This choice is analogous to assume a randomly oriented ambient magnetic 
   field which is capable to strongly limit the effects of the thermal 
   conduction. This assumption may affect some of the details of the 
   shock-cloud interaction especially in regions where the effects of 
   thermal conduction can be relevant. However we do not expect that 
   our assumption affects the large-scale structure of the remnant and 
   the ejecta distribution and, therefore, that it changes the main 
   conclusions of the paper.
   
   The morphology and the distribution of material observed in SNRs reflect 
   the interaction of the SN blast wave with the ambient environment
   as well as the physical processes associated to the SN explosion and 
   the internal structure of the progenitor star. Here we considered a 
   symmetric explosion for the initial conditions, which means that all 
   the features observed in our model arise from the interaction of the 
   remnant with the CSM/ISM. Thus, a correct description of the environment
   where the progenitor exploded has allowed us to explain the physical 
   origin of the puzzling and multi-thermal X-ray emission of IC~443. 
   The centrally peaked morphology (characteristic of MMSNRs) is a natural 
   result of the interaction with the complex environment of IC~443. 
   A combination of high resolution X-ray observations and accurate 3D HD 
   modelling is necessary to confirm if this scenario is applicable to other 
   MMSNRs. Our model also provided tight constraints on the explosion energy 
   and the remnant age, and supported the association of IC~443 with PWN 
   CXOU~J061705.3+222127. A more detailed description of the SN explosion 
   in IC~443 could add new information on the distribution of chemical 
   abundances within the ejecta and on intrinsic inhomogeneities like the 
   overionized jet-like structure described by \cite{gre18}.

\begin{acknowledgements}

   We thank the referee for useful comments and suggestions that 
   allowed us to improve the manuscript.
   We acknowledge the CINECA ISCRA initiative (Award HP10CET086), 
   the MoU INAF-CINECA initiative (Grant INA17\_C5A43) and the HPC 
   facility (SCAN) of the INAF – Osservatorio Astronomico di Palermo 
   for the availability of high performance computing resources and support.
   The PLUTO code, used in this work, was developed at the Turin 
   Astronomical Observatory in collaboration with the Department of 
   General Physics of Turin University and the SCAI Department of CINECA. 
   SO, MM, FB acknowledge financial contribution from the INAF mainstream 
   program and from the agreement ASI-INAF n.2017-14-H.O.
   We are grateful to Bob Franke, Focal Pointe Observatory, for 
   sharing the optical image of SNR~IC~443 used in the online movie and 
   the 3D interactive graphic.

\end{acknowledgements}

%
\bibliographystyle{aa} 
\bibliography{biblio.bib} 

\begin{thebibliography}{60}
\expandafter\ifx\csname natexlab\endcsname\relax\def\natexlab#1{#1}\fi

\bibitem[{{Abdo} {et~al.}(2010){Abdo}, {Ackermann}, {Ajello}, {Baldini},
  {Ballet}, {Barbiellini}, {Bastieri}, {Baughman}, {Bechtol}, {Bellazzini},
  {Berenji}, {Blandford}, {Bloom}, {Bonamente}, {Borgland}, {Bregeon}, {Brez},
  {Brigida}, {Bruel}, {Burnett}, {Buson}, {Caliandro}, {Cameron}, {Caraveo},
  {Casand jian}, {Cecchi}, {{c{C}}elik}, {Chekhtman}, {Cheung}, {Chiang},
  {Cillis}, {Ciprini}, {Claus}, {Cohen-Tanugi}, {Cominsky}, {Conrad}, {Cutini},
  {Dermer}, {de Angelis}, {de Palma}, {Silva}, {Drell}, {Drlica-Wagner},
  {Dubois}, {Dumora}, {Farnier}, {Favuzzi}, {Fegan}, {Focke}, {Fortin},
  {Frailis}, {Fukazawa}, {Funk}, {Fusco}, {Gargano}, {Gasparrini}, {Gehrels},
  {Germani}, {Giavitto}, {Giebels}, {Giglietto}, {Giordano}, {Glanzman},
  {Godfrey}, {Grenier}, {Grondin}, {Grove}, {Guillemot}, {Guiriec}, {Hanabata},
  {Harding}, {Hayashida}, {Hughes}, {Jackson}, {J{'o}hannesson}, {Johnson},
  {Johnson}, {Johnson}, {Kamae}, {Katagiri}, {Kataoka}, {Kawai}, {Kerr},
  {Kn{"o}dlseder}, {Kocian}, {Kuss}, {Land e}, {Latronico}, {Lee},
  {Lemoine-Goumard}, {Longo}, {Loparco}, {Lott}, {Lovellette}, {Lubrano},
  {Madejski}, {Makeev}, {Mazziotta}, {Meurer}, {Michelson}, {Mitthumsiri},
  {Moiseev}, {Monte}, {Monzani}, {Morselli}, {Moskalenko}, {Murgia},
  {Nakamori}, {Nolan}, {Norris}, {Nuss}, {Ohsugi}, {Orlando}, {Ormes}, {Ozaki},
  {Paneque}, {Panetta}, {Parent}, {Pelassa}, {Pepe}, {Pesce-Rollins}, {Piron},
  {Porter}, {Rain{`o}}, {Rando}, {Razzano}, {Reimer}, {Reimer}, {Reposeur},
  {Rochester}, {Rodriguez}, {Romani}, {Roth}, {Ryde}, {Sadrozinski}, {Sanchez},
  {Sander}, {Saz Parkinson}, {Scargle}, {Sgr{`o}}, {Siskind}, {Smith}, {Smith},
  {Spandre}, {Spinelli}, {Strickman}, {Strong}, {Suson}, {Tajima}, {Takahashi},
  {Takahashi}, {Tanaka}, {Thayer}, {Thayer}, {Thompson}, {Tibaldo}, {Torres},
  {Tosti}, {Tramacere}, {Uchiyama}, {Usher}, {Van Etten}, {Vasileiou},
  {Venter}, {Vilchez}, {Vitale}, {Waite}, {Wang}, {Winer}, {Wood}, {Ylinen}, \&
  {Ziegler}}]{abd10}
{Abdo}, A.~A., {Ackermann}, M., {Ajello}, M., {et~al.} 2010, \apj, 712, 459

\bibitem[{{Arnaud}(1996)}]{arn96}
{Arnaud}, K.~A. 1996, in Astronomical Society of the Pacific Conference Series,
  Vol. 101, Astronomical Data Analysis Software and Systems V, ed. G.~H.
  {Jacoby} \& J.~{Barnes}, 17

\bibitem[{{Bocchino} \& {Bykov}(2000)}]{boc00}
{Bocchino}, F. \& {Bykov}, A.~M. 2000, \aap, 362, L29

\bibitem[{{Bocchino} \& {Bykov}(2003)}]{boc03}
{Bocchino}, F. \& {Bykov}, A.~M. 2003, \aap, 400, 203

\bibitem[{{Bocchino} {et~al.}(2008){Bocchino}, {Krassilchtchikov},
  {Kretschmar}, {Bykov}, {Uvarov}, \& {Osipov}}]{boc08}
{Bocchino}, F., {Krassilchtchikov}, A.~M., {Kretschmar}, P., {et~al.} 2008,
  Adv. Space Res., 41, 396

\bibitem[{{Bocchino} {et~al.}(2009){Bocchino}, {Miceli}, \& {Troja}}]{boc09}
{Bocchino}, F., {Miceli}, M., \& {Troja}, E. 2009, \aap, 498, 139

\bibitem[{{Bocchino} {et~al.}(2001){Bocchino}, {Parmar}, {Mereghetti}, {Orland
  ini}, {Santangelo}, \& {Angelini}}]{boc01}
{Bocchino}, F., {Parmar}, A.~N., {Mereghetti}, S., {et~al.} 2001, \aap, 367,
  629

\bibitem[{{Braun} \& {Strom}(1986)}]{bra86}
{Braun}, R. \& {Strom}, R.~G. 1986, \aap, 164, 193

\bibitem[{{Burton} {et~al.}(1988){Burton}, {Geballe}, {Brand}, \&
  {Webster}}]{bur88}
{Burton}, M.~G., {Geballe}, T.~R., {Brand}, P.~W.~J.~L., \& {Webster}, A.~S.
  1988, \mnras, 231, 617

\bibitem[{{Bykov} {et~al.}(2008){Bykov}, {Krassilchtchikov}, {Uvarov},
  {Bloemen}, {Bocchino}, {Dubner}, {Giacani}, \& {Pavlov}}]{byk08}
{Bykov}, A.~M., {Krassilchtchikov}, A.~M., {Uvarov}, Y.~A., {et~al.} 2008,
  \apj, 676, 1050

\bibitem[{{Chevalier}(1999)}]{che99}
{Chevalier}, R.~A. 1999, \apj, 511, 798

\bibitem[{{Chevalier}(2005)}]{che05}
{Chevalier}, R.~A. 2005, \apj, 619, 839

\bibitem[{{Colella} \& {Woodward}(1984)}]{col84}
{Colella}, P. \& {Woodward}, P.~R. 1984, J. Comput. Phys., 54, 174

\bibitem[{{Cornett} {et~al.}(1977){Cornett}, {Chin}, \& {Knapp}}]{cor77}
{Cornett}, R.~H., {Chin}, G., \& {Knapp}, G.~R. 1977, \aap, 54, 889

\bibitem[{{Cox} {et~al.}(1999){Cox}, {Shelton}, {Maciejewski}, {Smith},
  {Plewa}, {Pawl}, \& {R{\'o}{\.z}yczka}}]{cox99}
{Cox}, D.~P., {Shelton}, R.~L., {Maciejewski}, W., {et~al.} 1999, \apj, 524,
  179

\bibitem[{{Denoyer}(1978)}]{den78}
{Denoyer}, L.~K. 1978, \mnras, 183, 187

\bibitem[{{Drake} \& {Orlando}(2010)}]{dra10}
{Drake}, J.~J. \& {Orlando}, S. 2010, \apjl, 720, L195

\bibitem[{{Gaensler} {et~al.}(2006){Gaensler}, {Chatterjee}, {Slane}, {van der
  Swaluw}, {Camilo}, \& {Hughes}}]{gae06}
{Gaensler}, B.~M., {Chatterjee}, S., {Slane}, P.~O., {et~al.} 2006, \apj, 648,
  1037

\bibitem[{{Ghavamian} {et~al.}(2007){Ghavamian}, {Laming}, \&
  {Rakowski}}]{gha07}
{Ghavamian}, P., {Laming}, J.~M., \& {Rakowski}, C.~E. 2007, \apjl, 654, L69

\bibitem[{{Greco} {et~al.}(2018){Greco}, {Miceli}, {Orlando}, {Peres}, {Troja},
  \& {Bocchino}}]{gre18}
{Greco}, E., {Miceli}, M., {Orlando}, S., {et~al.} 2018, \aap, 615, A157

\bibitem[{{Greco} {et~al.}(2020){Greco}, {Vink}, {Miceli}, {Orland o},
  {Dom{\v{c}}ek}, {Zhou}, {Bocchino}, \& {Peres}}]{gre20}
{Greco}, E., {Vink}, J., {Miceli}, M., {et~al.} 2020, \aap, 638, A101

\bibitem[{{Kashyap} \& {Drake}(2000)}]{kas00}
{Kashyap}, V. \& {Drake}, J.~J. 2000, Bull. Astron. Soc. India, 28, 475

\bibitem[{{Lazarian}(2006)}]{laz06}
{Lazarian}, A. 2006, \apjl, 645, L25

\bibitem[{{Leahy}(2004)}]{lea04}
{Leahy}, D.~A. 2004, \aj, 127, 2277

\bibitem[{{Lee} {et~al.}(2012){Lee}, {Koo}, {Snell}, {Yun}, {Heyer}, \&
  {Burton}}]{lee12}
{Lee}, J.-J., {Koo}, B.-C., {Snell}, R.~L., {et~al.} 2012, \apj, 749, 34

\bibitem[{{Lee} {et~al.}(2008){Lee}, {Koo}, {Yun}, {Stanimirovi{'c}}, {Heiles},
  \& {Heyer}}]{lee08}
{Lee}, J.-J., {Koo}, B.-C., {Yun}, M.~S., {et~al.} 2008, \aj, 135, 796

\bibitem[{{Matsumura} {et~al.}(2017){Matsumura}, {Tanaka}, {Uchida}, {Okon}, \&
  {Tsuru}}]{mat17}
{Matsumura}, H., {Tanaka}, T., {Uchida}, H., {Okon}, H., \& {Tsuru}, T.~G.
  2017, \apj, 851, 73

\bibitem[{{Matzner} \& {McKee}(1999)}]{mat99}
{Matzner}, C.~D. \& {McKee}, C.~F. 1999, \apj, 510, 379

\bibitem[{{Miceli} {et~al.}(2010){Miceli}, {Bocchino}, {Decourchelle},
  {Ballet}, \& {Reale}}]{mic10}
{Miceli}, M., {Bocchino}, F., {Decourchelle}, A., {Ballet}, J., \& {Reale}, F.
  2010, \aap, 514, L2

\bibitem[{{Mignone} {et~al.}(2007){Mignone}, {Bodo}, {Massaglia}, {Matsakos},
  {Tesileanu}, {Zanni}, \& {Ferrari}}]{mig07}
{Mignone}, A., {Bodo}, G., {Massaglia}, S., {et~al.} 2007, \apjs, 170, 228

\bibitem[{{Miller} \& {Colella}(2002)}]{mil02}
{Miller}, G.~H. \& {Colella}, P. 2002, J. Comput. Phys., 183, 26

\bibitem[{{Narayan} \& {Medvedev}(2001)}]{nar01}
{Narayan}, R. \& {Medvedev}, M.~V. 2001, \apjl, 562, L129

\bibitem[{{Okon} {et~al.}(2020){Okon}, {Tanaka}, {Uchida}, {Yamaguchi},
  {Tsuru}, {Seta}, {Smith}, {Yoshiike}, {Orlando}, {Bocchino}, \&
  {Miceli}}]{oko20}
{Okon}, H., {Tanaka}, T., {Uchida}, H., {et~al.} 2020, \apj, 890, 62

\bibitem[{{Olbert} {et~al.}(2001){Olbert}, {Clearfield}, {Williams}, {Keohane},
  \& {Frail}}]{olb01}
{Olbert}, C.~M., {Clearfield}, C.~R., {Williams}, N.~E., {Keohane}, J.~W., \&
  {Frail}, D.~A. 2001, \apjl, 554, L205

\bibitem[{{Orlando} {et~al.}(2008){Orlando}, {Bocchino}, {Reale}, {Peres}, \&
  {Pagano}}]{orl08}
{Orlando}, S., {Bocchino}, F., {Reale}, F., {Peres}, G., \& {Pagano}, P. 2008,
  \apj, 678, 274

\bibitem[{{Orlando} {et~al.}(2009){Orlando}, {Drake}, \& {Laming}}]{orl09}
{Orlando}, S., {Drake}, J.~J., \& {Laming}, J.~M. 2009, \aap, 493, 1049

\bibitem[{{Orlando} {et~al.}(2019){Orlando}, {Miceli}, {Petruk}, {Ono},
  {Nagataki}, {Aloy}, {Mimica}, {Lee}, {Bocchino}, {Peres}, \&
  {Guarrasi}}]{orl19}
{Orlando}, S., {Miceli}, M., {Petruk}, O., {et~al.} 2019, \aap, 622, A73

\bibitem[{{Orlando} {et~al.}(2015){Orlando}, {Miceli}, {Pumo}, \&
  {Bocchino}}]{orl15}
{Orlando}, S., {Miceli}, M., {Pumo}, M.~L., \& {Bocchino}, F. 2015, \apj, 810,
  168

\bibitem[{{Orlando} {et~al.}(2005){Orlando}, {Peres}, {Reale}, {Bocchino},
  {Rosner}, {Plewa}, \& {Siegel}}]{orl05}
{Orlando}, S., {Peres}, G., {Reale}, F., {et~al.} 2005, \aap, 444, 505

\bibitem[{{Petre} {et~al.}(1988){Petre}, {Szymkowiak}, {Seward}, \&
  {Willingale}}]{pet88}
{Petre}, R., {Szymkowiak}, A.~E., {Seward}, F.~D., \& {Willingale}, R. 1988,
  \apj, 335, 215

\bibitem[{{Petruk}(2001)}]{pet01}
{Petruk}, O. 2001, \aap, 371, 267

\bibitem[{{Rho} {et~al.}(2001){Rho}, {Jarrett}, {Cutri}, \& {Reach}}]{rho01}
{Rho}, J., {Jarrett}, T.~H., {Cutri}, R.~M., \& {Reach}, W.~T. 2001, \apj, 547,
  885

\bibitem[{{Rho} \& {Petre}(1998)}]{rho98}
{Rho}, J. \& {Petre}, R. 1998, \apjl, 503, L167

\bibitem[{{Shelton} {et~al.}(1999){Shelton}, {Cox}, {Maciejewski}, {Smith},
  {Plewa}, {Pawl}, \& {R{\'o}{\.z}yczka}}]{she99}
{Shelton}, R.~L., {Cox}, D.~P., {Maciejewski}, W., {et~al.} 1999, \apj, 524,
  192

\bibitem[{{Shinn} {et~al.}(2011){Shinn}, {Koo}, {Seon}, \& {Lee}}]{shi11}
{Shinn}, J.-H., {Koo}, B.-C., {Seon}, K.-I., \& {Lee}, H.-G. 2011, \apj, 732,
  124

\bibitem[{{Slane} {et~al.}(2015){Slane}, {Bykov}, {Ellison}, {Dubner}, \&
  {Castro}}]{sla15}
{Slane}, P., {Bykov}, A., {Ellison}, D.~C., {Dubner}, G., \& {Castro}, D. 2015,
  \ssr, 188, 187

\bibitem[{{Smith} {et~al.}(2001){Smith}, {Brickhouse}, {Liedahl}, \&
  {Raymond}}]{smi01}
{Smith}, R.~K., {Brickhouse}, N.~S., {Liedahl}, D.~A., \& {Raymond}, J.~C.
  2001, \apjl, 556, L91

\bibitem[{{Snell} {et~al.}(2005){Snell}, {Hollenbach}, {Howe}, {Neufeld},
  {Kaufman}, {Melnick}, {Bergin}, \& {Wang}}]{sne05}
{Snell}, R.~L., {Hollenbach}, D., {Howe}, J.~E., {et~al.} 2005, \apj, 620, 758

\bibitem[{{Su} {et~al.}(2014){Su}, {Fang}, {Yang}, {Zhou}, \& {Chen}}]{su14}
{Su}, Y., {Fang}, M., {Yang}, J., {Zhou}, P., \& {Chen}, Y. 2014, \apj, 788,
  122

\bibitem[{{Swartz} {et~al.}(2015){Swartz}, {Pavlov}, {Clarke}, {Castelletti},
  {Zavlin}, {Bucciantini}, {Karovska}, {van der Horst}, {Yukita}, \&
  {Weisskopf}}]{swa15}
{Swartz}, D.~A., {Pavlov}, G.~G., {Clarke}, T., {et~al.} 2015, \apj, 808, 84

\bibitem[{{Tavani} {et~al.}(2010){Tavani}, {Giuliani}, {Chen}, {Argan},
  {Barbiellini}, {Bulgarelli}, {Caraveo}, {Cattaneo}, {Cocco}, {Contessi},
  {D'Ammand o}, {Costa}, {De Paris}, {Del Monte}, {Di Cocco}, {Donnarumma},
  {Evangelista}, {Ferrari}, {Feroci}, {Fuschino}, {Galli}, {Gianotti},
  {Labanti}, {Lapshov}, {Lazzarotto}, {Lipari}, {Longo}, {Marisaldi},
  {Mastropietro}, {Mereghetti}, {Morelli}, {Moretti}, {Morselli}, {Pacciani},
  {Pellizzoni}, {Perotti}, {Piano}, {Picozza}, {Pilia}, {Pucella}, {Prest},
  {Rapisarda}, {Rappoldi}, {Scalise}, {Rubini}, {Sabatini}, {Striani},
  {Soffitta}, {Trifoglio}, {Trois}, {Vallazza}, {Vercellone}, {Vittorini},
  {Zambra}, {Zanello}, {Pittori}, {Verrecchia}, {Santolamazza}, {Giommi},
  {Colafrancesco}, {Antonelli}, \& {Salotti}}]{tav10}
{Tavani}, M., {Giuliani}, A., {Chen}, A.~W., {et~al.} 2010, \apjl, 710, L151

\bibitem[{{Troja} {et~al.}(2008){Troja}, {Bocchino}, {Miceli}, \&
  {Reale}}]{tro08}
{Troja}, E., {Bocchino}, F., {Miceli}, M., \& {Reale}, F. 2008, \aap, 485, 777

\bibitem[{{Troja} {et~al.}(2006){Troja}, {Bocchino}, \& {Reale}}]{tro06}
{Troja}, E., {Bocchino}, F., \& {Reale}, F. 2006, \apj, 649, 258

\bibitem[{{Welsh} \& {Sallmen}(2003)}]{wel03}
{Welsh}, B.~Y. \& {Sallmen}, S. 2003, \aap, 408, 545

\bibitem[{{White} \& {Long}(1991)}]{whi91}
{White}, R.~L. \& {Long}, K.~S. 1991, \apj, 373, 543

\bibitem[{{Yamaguchi} {et~al.}(2009){Yamaguchi}, {Ozawa}, {Koyama}, {Masai},
  {Hiraga}, {Ozaki}, \& {Yonetoku}}]{yam09}
{Yamaguchi}, H., {Ozawa}, M., {Koyama}, K., {et~al.} 2009, \apjl, 705, L6

\bibitem[{{Yamaguchi} {et~al.}(2018){Yamaguchi}, {Tanaka}, {Wik}, {Rho},
  {Bamba}, {Castro}, {Smith}, {Foster}, {Uchida}, {Petre}, \&
  {Williams}}]{yam18}
{Yamaguchi}, H., {Tanaka}, T., {Wik}, D.~R., {et~al.} 2018, \apjl, 868, L35

\bibitem[{{Zhang} \& {Chevalier}(2019)}]{zha19}
{Zhang}, D. \& {Chevalier}, R.~A. 2019, \mnras, 482, 1602

\bibitem[{{Zhang} {et~al.}(2018){Zhang}, {Tang}, {Zhang}, {Sun}, {Gotthelf},
  {Zhang}, {Li}, {Cheng}, {Pasham}, {Baganoff}, {Perez}, {Hailey}, \&
  {Mori}}]{zha18}
{Zhang}, S., {Tang}, X., {Zhang}, X., {et~al.} 2018, \apj, 859, 141

\bibitem[{{Zhou} {et~al.}(2011){Zhou}, {Miceli}, {Bocchino}, {Orland o}, \&
  {Chen}}]{zho11}
{Zhou}, X., {Miceli}, M., {Bocchino}, F., {Orland o}, S., \& {Chen}, Y. 2011,
  \mnras, 415, 244

\end{thebibliography}
%

\begin{appendix}

\section{SN explosion at the geometrical center of the atomic and molecular clouds}
\label{app:sim}

   \begin{figure*}[!ht]
      \includegraphics*[width=\hsize]{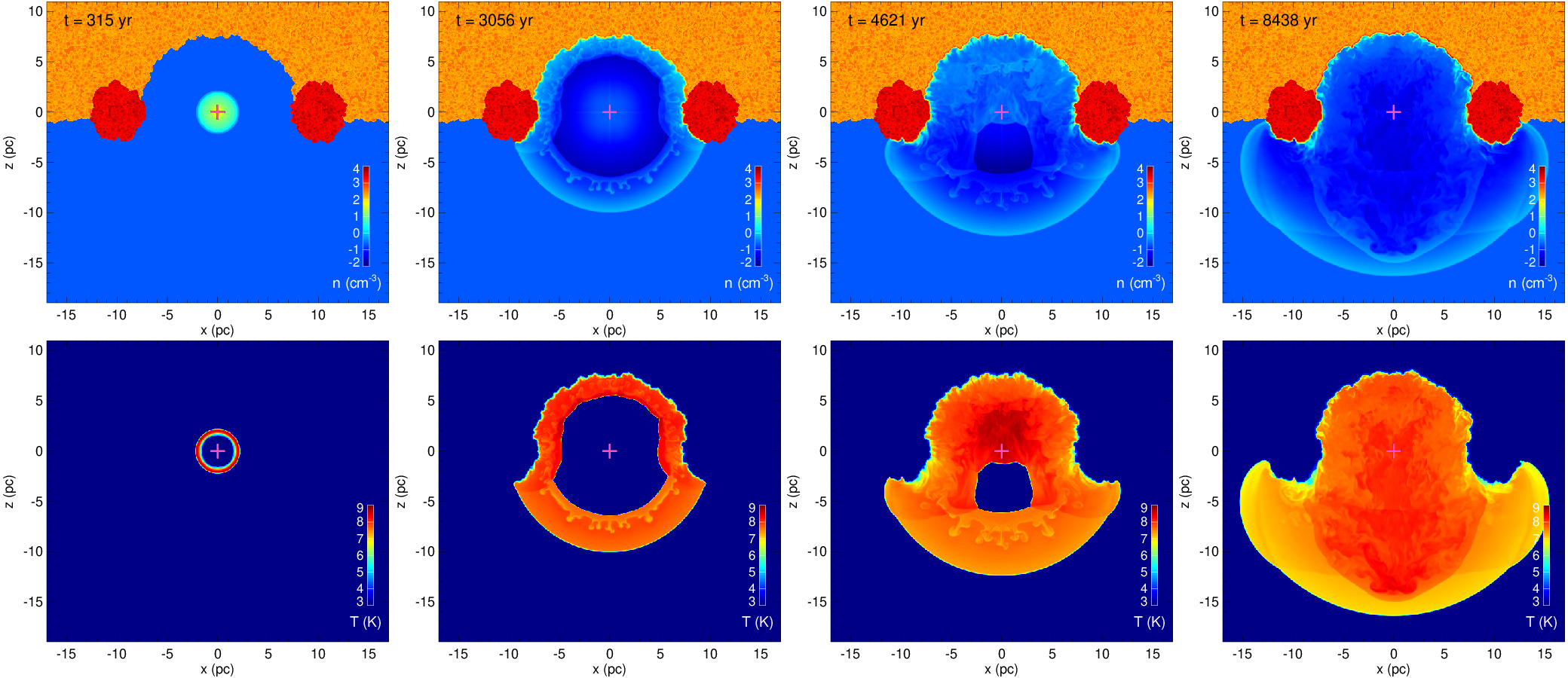}
      \includegraphics*[width=\hsize]{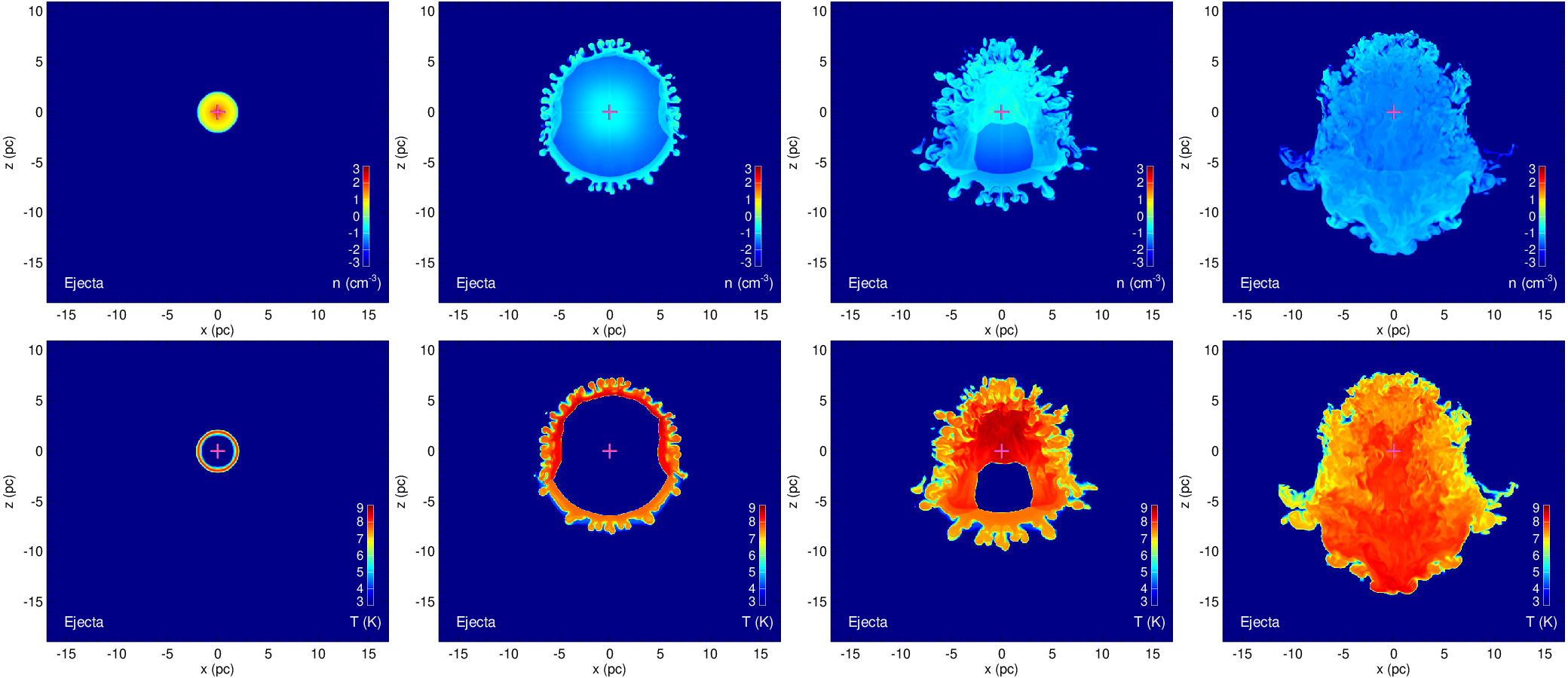}
      \caption{Density (first row; third row, only the ejecta) and ionic 
      temperature (second row; fourth row, only the ejecta) 
      distributions in logarithmic scale in the $(x,0,z)$ plane at different 
      evolution times (increasing from left to right).
      The magenta cross indicates the position of the explosion.}
      \label{Fig:DensTempC}%
   \end{figure*}
   
   \begin{figure*}
      \includegraphics*[width=\hsize]{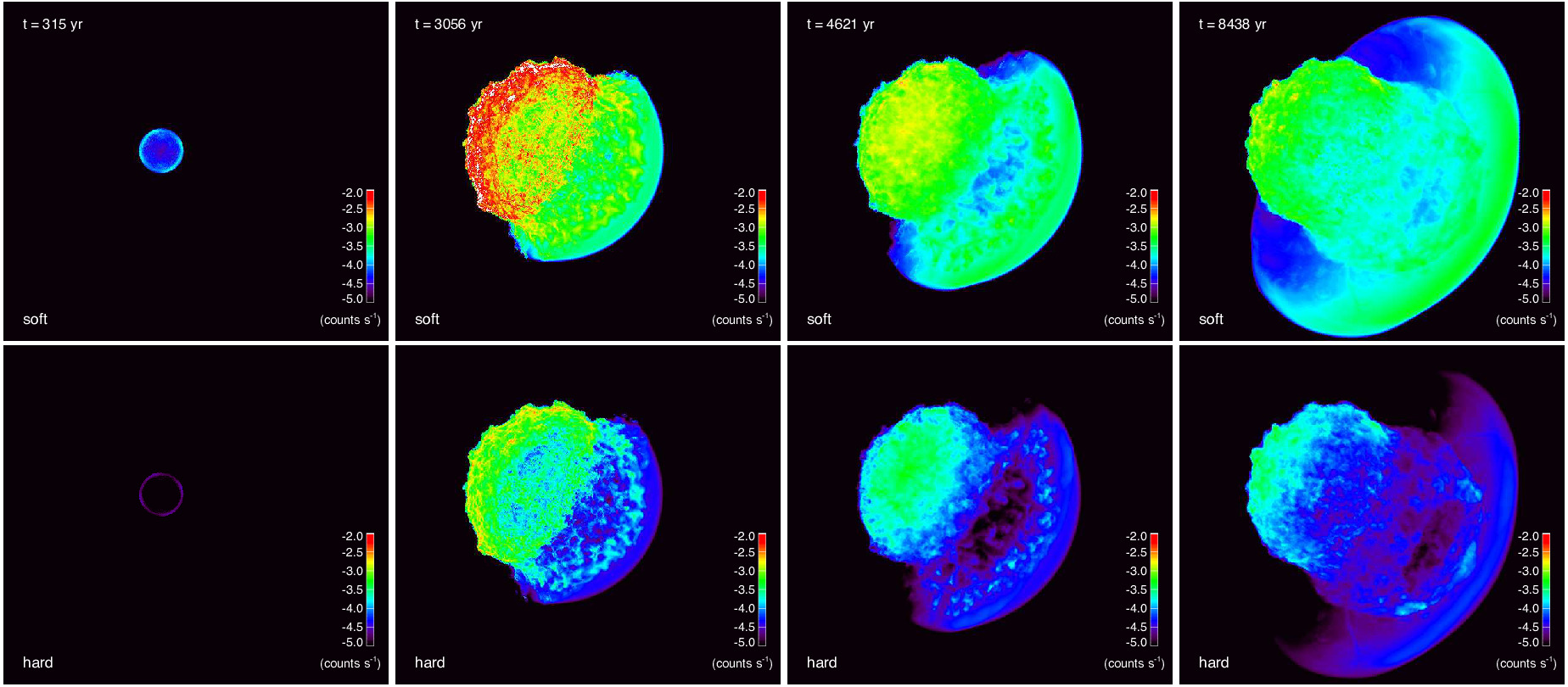}
      \caption{Synthetic X-ray count rate maps in the [$0.5-1.4$]~keV band 
      (upper panels) and [$1.4-5$]~keV band (bottom panels) in logarithmic 
      scale, at different evolution times (increasing from left to right).}
      \label{Fig:XrayC}%
   \end{figure*}

   In this section we present the case with $d_\mathrm{x}=0$, which 
   corresponds to an explosion centered in the cavity formed by the 
   surrounding clouds (see Fig.~\ref{Fig:Initial} and Sec.~\ref{sec:num}). 
   The rest of the parameters are the same as in our favourite model 
   (see Table~\ref{parameters}).
   
   The evolution of the model is shown in Fig.~\ref{Fig:DensTempC} which, 
   similarly to Fig.~\ref{Fig:DensTemp}, reports the density and temperature 
   distributions in the $[x,z]$ plane, using same scales and evolution times 
   for comparison with the best model described in Sec.~\ref{sec:results}. 
   The ejecta expanding through the uniform intercloud medium start to 
   interact with the atomic and molecular clouds $\sim 2000$~yr after the 
   SN event, much later than in the best case (see Fig.~\ref{Fig:DensTemp}, 
   and Fig.~\ref{Fig:DensTempC}) and the forward shock hits the clouds at the 
   same time in all directions.
   As a consequence the reverse shock (powered by the interaction with the 
   clouds) travels through the ejecta when already expanded and partially 
   refocus slightly southern to the origin of the explosion indicated with 
   a magenta cross (see second and third columns in Fig.~\ref{Fig:DensTempC}). 
   As a result of this evolution the remnant becomes symmetric respect to 
   the $z$-axis (see last column in Fig.~\ref{Fig:DensTempC}). 
   
   In Figure~\ref{Fig:XrayC} we present X-ray count rate maps synthesized 
   (see Sec.~\ref{sec:synthesis}) from the model at the same evolution times 
   as in Fig.~\ref{Fig:DensTempC} (increasing from left to right), considering 
   the LoS as in our best model (see Fig.~\ref{Fig:Ejecta}, left panels).
   The X-ray emission maps of the remnant show a symmetric distribution 
   respect to the axis of the toroidal structure (see Fig.\ref{Fig:Initial} 
   and Fig.~\ref{Fig:Ejecta}) due to the symmetric interaction with the clouds 
   (see Fig.~\ref{Fig:XrayC}). This distribution does not agree with the 
   X-ray observations \citep{tro06,tro08,boc09,gre18} and do not reproduce 
   the X-ray morphology of IC~443 (see left panels in Fig.\ref{Fig:XrayObsMod}).

\section{\textit{XMM-Newton} observations of IC~443}
\label{app:obs}

   The source IC~443 has been observed several times with the 
   \textit{European Photon Imaging Camera} (EPIC) on board of 
   \textit{XMM-Newton}. We carried out spectral analysis of the 
   observation performed in March 2010 (Obs-ID = 0600110101, PI: E. Troja)
   by considering only MOS2 data. For the aim of this work, only one camera 
   is needed and, out of the three available, we chose MOS2 since it is the 
   less degraded.
   We built count-rate images in two different energy bands following 
   the same approach described in Sect. 2 in \cite{gre18}, considering 
   a bin size of 11\arcsec.
   We used the \textit{Science Analysis System} (SAS), version 16.1.0, 
   to perform the whole data analysis. In particular, we used the SAS 
   tool evigweight to correct vignetting effect in the spectra; we applied 
   the SAS tasks rmfgen and arfgen obtaining response and ancillary 
   matrices; and we binned spectra to obtain at least 25 counts per bin. 
   The spectral analysis has been performed with XSPEC 
   (version 12.10.0c, \citealt{arn96}).

\section{Online material}
\label{app:online}

\begin{itemize}
\setlength\itemsep{0.5em}
 
   \item \textbf{Movie~1}: Temporal evolution of the density (first column) 
   and the ionic temperature (second column) distributions in logarithmic 
   scale in the $(x,0,z)$ plane. The second row shows the distributions 
   considering only the ejecta. The magenta cross indicates the position of 
   the explosion.
    
   \item  \textbf{Movie~2}: Isosurface of the distribution of density at 
   $t\approx 8400$~yr for the ejecta of the favourite model for SNR~IC~443.
   The opaque irregular isosurface corresponds to a value of density which 
   is at 1\% of the peak density; their colors give the radial velocity 
   in units of 1000~km~s$^{-1}$ on the isosurface.
   The semi-transparent surface marks the position of the forward shock; the 
   toroidal semi-transparent structure in purple represents the molecular cloud.
   A navigable 3D graphic is available at \url{https://skfb.ly/6W9oM}.

   \item  \textbf{Movie~3}: Isosurface of the distribution of density at 
   $t\approx 8400$~yr for the ejecta of the favourite model for SNR~IC~443.
   The opaque irregular isosurface corresponds to a value of density which 
   is at 5\% of the peak density; their colors give the radial velocity 
   in units of 1000~km~s$^{-1}$ on the isosurface.
   The semi-transparent surface marks the position of the forward shock; 
   the toroidal semi-transparent structure in purple represents the 
   molecular cloud.

   \item  \textbf{Movie~4}: Same as Movie~2, compared with an optical 
   observation of SNR~IC~443. The transparent image passing through the 
   center of the remnant is a wide field optical observation of the remnant.
   A navigable 3D graphic is available at \url{https://skfb.ly/6X6BV}.
   Image credit: Wide Field Optical: Bob Franke (Focal Pointe Observatory). 
   
   \item  \textbf{Movie~5}: Temporal evolution of the synthetic X-ray count 
   rate maps in the [$0.5-1.4$]~keV band (left panel) and [$1.4-5$]~keV band 
   (right panel) in logarithmic scale.

\end{itemize}

\end{appendix}

\end{document}